\documentclass[10pt]{article}
\usepackage{latexsym} \usepackage{epsf}
\usepackage{a4}
\usepackage{amsfonts}
\renewcommand{\theequation}{\thesection.\arabic{equation}}
\newcounter{subequation}[equation]
\makeatletter

\expandafter\let\expandafter\reset@font\csname reset@font\endcsname

\def\subeqnarray{\arraycolsep1pt
    \def\@eqnnum\stepcounter##1{\stepcounter{subequation}%
        {\reset@font\rm(\theequation\alph{subequation})}}
\jot5mm     \eqnarray}

\makeatother

\def\tr{\mathop{\hbox{\rm tr}}\nolimits}

\def\be{\begin{equation}}
\def\ee{\end{equation}}
\def\bea{\begin{eqnarray}}
\def\eea{\end{eqnarray}}
\def\ba{\begin{array}}
\def\ea{\end{array}}
\def\dd{\partial}

\def\one#1{#1^{\raise5pt\hbox{$\scriptstyle\!\!\!\!1$}}\,{}}
\def\two#1{#1^{\raise5pt\hbox{$\scriptstyle\!\!\!\!2$}}\,{}}

\def\II{\hbox{{1}\kern-.25em\hbox{l}}}
\makeatletter
\def\binrel@#1{\begingroup
  \setboxz@h{\thinmuskip0mu
    \medmuskip\m@ne mu\thickmuskip\@ne mu
    \setbox\tw@\hbox{$#1\m@th$}\kern-\wd\tw@
    ${}#1{}\m@th$}%
  \edef\@tempa{\endgroup\let\noexpand\binrel@@
    \ifdim\wdz@<\z@ \mathbin
    \else\ifdim\wdz@>\z@ \mathrel
    \else \relax\fi\fi}%
  \@tempa
}
\let\binrel@@\relax
\def\overset#1#2{\binrel@{#2}%
  \binrel@@{\mathop{\kern\z@#2}\limits^{#1}}}
\def\underset#1#2{\binrel@{#2}%
  \binrel@@{\mathop{\kern\z@#2}\limits_{#1}}}
\makeatother
\newfont{\bbd}{msbm10 scaled\magstep1}
\def\C{\hbox{\bbd C}}

\def\R{\hbox{\bbd R}}

\def\P{\hbox{\bbd P}}

\def\RR{{\mathcal R}}

\def\RR{{\mathcal R}}

\newtheorem{prop}{Proposition}

\begin{document}

{\begin{center}
{\LARGE {Baxter Q-operators of the XXZ chain \\
and R-matrix factorization} } \\ [8mm] {\large S.
Derkachov$^{a}$,
\footnote{e-mail:S.Derkachov@pobox.spbu.ru}, D.
Karakhanyan$^b$\footnote{e-mail: karakhan@lx2.yerphi.am} \&
R. Kirschner$^c$\footnote{e-mail:Roland.Kirschner@itp.uni-leipzig.de} \\
[3mm] }
\end{center}

\begin{itemize}
\item[$^a$]
Department of Mathematics, St Petersburg Technology Institute, \\
Sankt Petersburg, Russia
\item[$^b$]
Yerevan Physics Institute, \\
Br.Alikhanian st. 2, 375036, Yerevan, Armenia.
\item[$^c$] Institut f\"{u}r Theoretische Physik, Universit\"{a}t Leipzig, \\
PF 100920, D-04009 Leipzig, Germany
\end{itemize}

\vspace{3cm}

\begin{center}
{\bf Abstract}
\end{center}
\noindent 
We construct Baxter operators as generalized
transfer matrices being traces of products of generic
$R$ matrices. The latter are shown to factorize into simpler
operators allowing for explicit expressions in terms of
functions of a Weyl pair of basic operators. These explicit
expressions are the basis for explicit expression for
Baxter Q-operators and for investigating their properties.

\renewcommand{\refname}{References.}
\renewcommand{\thefootnote}{\arabic{footnote}}
\setcounter{footnote}{0}
\setcounter{equation}{0}

\renewcommand{\theequation}{\thesection.\arabic{equation}}
\setcounter{equation}{0}

\section{Introduction}

We consider a periodic chain with integrable dynamics
carrying generic representations $ \ell_i$ of the
$q$-deformed $sl(2)$ algebra on their sites, $i=1, ...,N$.
The considered infinite-dimensional representations of
lowest weight type are realized in the space of
polynomials.

Chains with infinite-dimensional representations on their
sites play a role in the investigation of gauge field
theories, in the high energy limit of scattering \cite{L,
FKo} and in the renormalization of composite operators
\cite{Braun}. In these applications the polynomials can
appear as wave functions, where the variable is Fourier
conjugated to the parton light-cone momentum or as the
expression of the particular structure of multiplicatively
renormalized composite operators where the variable is the
light-cone position of the field operators. Motivated by
these applications the methods of integrable chains have
been reformulated and developed in a number of papers.

The conserved charges of the integrable chain can be
obtained by expansion of the transfer matrix $t(u)$ or
equally well of a operator $Q(u,\ell_0)$ being the natural
generalization of the transfer matrix, $ t(u+ \frac{1}{2} )
= Q(u, \ell_0 = -\frac{1}{2})$, and obeying a Baxter
relation with the transfer matrix (to be formulated below).

The concept of $Q$-operator was introduced by Baxter in
analyzing the eight-vertex model \cite{Baxter}. Using
the universal $R$-matrix for the $U_q(\hat{s\ell}_2)$ affine
algebra the $Q$-operator for quantum KdV model was
constructed in~\cite{BLZ}. The general algebraic scheme how
to derive the algebraic relations between different objects
of the Quantum Inverse Scattering Method
 was formulated in the paper of A.Antonov and
B.Feigin~\cite{AF}. The Baxter $Q$-operators have been constructed
for different models in the
papers~\cite{PG,KSS,P,KP,RW,Z,SD,DKM1,Man}.
The $Q$ operators for spin $\frac{1}{2}$ XXZ chains have been studied
in \cite{Korff} emphasizing the loop algebra representation theory and the
case of $q$ being a root of unity.
Chains with q-deformed principal series representations of $SL(2, \R)$
are studied in \cite{BT}.

The idea of constructing $Q$-operators as  generalized
transfer matrices from $R$-operators acting in general
representations is due to Sklyanin \cite{Skl1,Skl2} and the
first explicit construction with this idea appeared in
\cite{Volkov}.

A factorization of the $R$ matrix, acting on the tensor
product of generic representations of the algebra
$s\ell(2)$, has been established in \cite{SD2} by
considering the action of $\mathrm{R}$ in the
$\mathrm{R}\mathrm{L}\mathrm{L}$ relation as the
interchanging of representation and spectral parameter in
the Lax matrices and decomposing this action into two
steps. The defining relation, simplified for the factors
$\RR_{\pm}$, has been solved in terms of functions of the
Heisenberg operator pair $x, \partial$. Then this explicit
operator solution allows the explicit construction of
Baxter operators \cite{SD3}. Probably the first example of
such factorization of $R$-matrix and Baxter's $Q$-operator has been
obtained in the context of the chiral Potts
model~\cite{BS,Tar,BOU}.

In the present paper the analogous factorization of the
Yang-Baxter $R$ matrix is performed in the $q$-deformed
$sl(2)$ symmetric (XXZ) case. The $R$-operator for generic
representations is represented in terms of functions of a
Weyl pair of basic operators and Baxter operators are
constructed.

Previously in \cite{KKM} the Yang-Baxter $R$ matrix acting on
generic representations of the $q$-deformed $sl(2)$ has
been represented in explicite spectral and integral forms.
Similar to the classical papers by Jimbo \cite{Jimbo} the
defining $\mathrm{R}\mathrm{L}\mathrm{L}$ relation has been
written equivalently in terms of the intertwining relation
of the Drinfeld co-product type. This equivalent
intertwining form of the defining relations will be the
starting point of deriving the operator solutions for the
factors $\RR_{\pm}$ of the generic Yang-Baxter $\mathrm{R}$
matrix.

Also in \cite{KKM} polynomial eigenfunctions  of the
$R$ matrix are given explicitly.
A generalization of these functions
 build the integral kernel of ${R}$.
The functions appearing in the operator solution obtained
in the present paper are closely related to the polynomial
eigenfunctions of $R$. With particular substitutions of
their parameters these polynomials serve also as
eigenfunctions of the factors $\RR_{\pm}$.

Using the explicit operator results about ${R}$
matrices and their factors $\RR_\pm$ we construct
generalized transfer matrices and proof that they are
Baxter operators, i.e. obey Baxter relations in products
with the ordinary transfer matrix.  We consider closed
chains in general and then discuss specific features
appearing in the important special case of homogeneous
closed chains.

\section{The quantum algebra $U_q(s\ell_2)$ and
the general R-matrix}

\setcounter{equation}{0}

The quantum algebra $U_q(s\ell_2)$ has three generators
$\mathbf{S}\ ,\ \mathbf{S}_{\pm}$ with commutation relations
$$
\left[\mathbf{S},\mathbf{S}_{\pm}\right] = \pm
\mathbf{S}_{\pm} \ ,\
\left[\mathbf{S}_{+},\mathbf{S}_{-}\right] =
\left[2\mathbf{S}\right]_q
$$
where
$$
\left[\mathbf{x}\right]_q \equiv
\frac{q^{\mathbf{x}}-q^{-\mathbf{x}}}{q-q^{-1}}
$$
is the standard definition of q-numbers.

We shall use the representation
$\mathrm{V}_{\ell}$ of $U_q(s\ell_2)$ in the
infinite-dimensional space $\C[z]$ of polynomials in
the variable $z$ with the standard monomial basis
$\left\{z^k\right\}_{k=0}^{\infty}$ and lowest weight
vector $v_0 = 1$. The action of generators in
$\mathrm{V}_{\ell}$ is given by the first-order
differential operators: \be \mathrm{S} = z\dd + \ell \ ,\
\mathrm{S}_{-} =
-\frac{1}{z}\cdot\left[z\dd\right]_q\ ,\
\mathrm{S}_{+} = z\cdot \left[z\dd+2\ell\right]_q
\label{diff} \ee

The Lax-operator has the form~\cite{F,Bytsko,KKM}
$$
\mathrm{L}(u) \equiv
\left(\begin{array}{cc} \left[u+\mathrm{S}\right]_q
&
\mathrm{S}_{-} \\
\mathrm{S}_{+} &
\left[u-\mathrm{S}\right]_q
   \end{array}\right) = \left(\begin{array}{cc}
\left[u+\ell+z\dd\right]_q & -\frac{1}{z}\cdot
\left[z\dd\right]_q \\
z \cdot\left[2\ell+z\dd\right]_q &
\left[u-\ell-z\dd\right]_q
   \end{array}\right)
$$
The Lax operator acts in the space $\C[z]\otimes\C^2$ and,
despite the compact notation,  $\mathrm{L}(u)$ depends actually
on two parameters: spin $\ell$ and spectral parameter $u$.
We shall use the parametrization $u_{+}\equiv
u+\ell, u_{-}\equiv u-\ell$ and display all parameters
explicitly:
$$
\mathrm{L}(u_+,u_-) \equiv \left(\begin{array}{cc}
\left[u_+ +z\dd\right]_q & \mathrm{S}_{-} \\
z\cdot \left[u_+ -u_- +z\dd\right]_q &
\left[u_- -z\dd\right]_q
   \end{array}\right)
$$
except for the generator $\mathrm{S}_{-}$ because $\mathrm{S}_{-} =
-\frac{1}{z}\cdot\left[z\dd\right]_q$ does not depend on
parameters $u_{+}$ and $u_{-}$.

There exists a very useful factorized representation for the
$\mathrm{L}$-operator~\cite{KZ}
\begin{equation}
\mathrm{L}(u_+,u_-) = \frac{1}{q-q^{-1}}\cdot
\left(\begin{array}{cc}
1 & 1 \\
z q^{-u_-}& z q^{u_-}\end{array}\right)\
\left(\begin{array}{cc}
q^{z\dd} & 0 \\
0& q^{-z\dd}\end{array}\right)\ \left(\begin{array}{cc}
q^{u_+} & -\frac{q}{z} \\
-q^{-u_+} & \frac{1}{q z}
\end{array}\right)
\label{factor}
\end{equation}
The defining system of equations for the $\R$-operator has the form
\begin{equation}
\R_{12}(u-v)\cdot
\mathrm{L}_{1}(u_+,u_-)\mathrm{L}_{2}(v_+,v_-)=
\mathrm{L}_{2}(v_+,v_-)\mathrm{L}_{1}(u_+,u_-)
\cdot\R_{12}(u-v). \label{Df}
\end{equation}
It is useful to extract the operator of permutation
$\P_{12}$
$$
\P_{12}\Psi(z_1,z_2) = \Psi(z_2,z_1)\ ;\
\Psi(z_1,z_2)\in\C[z_1,z_2]
$$
from the $\R$-operator $\R_{12}(u-v) =
\P_{12}\check{\R}_{12}(u_{+},u_-|v_{+},v_-)$ and write the defining
equation for the $\check{\R}$-operator (the check form of YB relation)
\begin{equation}
\check{\R}_{12}(u_{+},u_-|v_{+},v_-)
\mathrm{L}_{1}(u_+,u_-)\mathrm{L}_{2}(v_+,v_-)=
\mathrm{L}_{1}(v_+,v_-)\mathrm{L}_{2}(u_+,u_-)
\check{\R}_{12}(u_{+},u_-|v_{+},v_-).
\label{DF}
\end{equation}
\begin{prop}
The defining system of equations~(\ref{DF}) for the
operator $\check{\R}_{12}$ is equivalent to the following
system of equations~\cite{Jimbo,KKM,Bytsko}
\begin{equation}
\check{\R}_{12} \cdot q^{z_1\dd_1+z_2\dd_2} =
q^{z_1\dd_1+z_2\dd_2} \cdot \check{\R}_{12}
\label{R1}
\end{equation}
\begin{equation}
\check{\R}_{12} \cdot \left(
\mathrm{S}^{-}_1\cdot
q^{\pm(v_--z_2\dd_2)} +
\mathrm{S}^{-}_2\cdot
q^{\pm(u_++z_1\dd_1)}\right) =
\left(
\mathrm{S}^{-}_1\cdot
q^{\pm(u_--z_2\dd_2)} +
\mathrm{S}^{-}_2\cdot
q^{\pm(v_++z_1\dd_1)}
\right)\cdot \check{\R}_{12}
\label{R2}
\end{equation}
$$
\check{\R}_{12} \cdot \left(
z_1\cdot\left[z_1\dd_1+u_+-u_-\right]_q
\cdot q^{\pm(v_++z_2\dd_2)} +
z_2\cdot\left[z_2\dd_2+v_+-v_-\right]_q
\cdot q^{\pm(u_--z_1\dd_1)} \right) =
$$
\begin{equation}
= \left(
z_1\cdot\left[z_1\dd_1+v_+-v_-\right]_q
\cdot q^{\pm(u_++z_2\dd_2)} +
z_2\cdot\left[z_2\dd_2+u_+-u_-\right]_q
\cdot q^{\pm(v_--z_1\dd_1)}
\right)\cdot \check{\R}_{12}
\label{R3}
\end{equation}
\end{prop}
We shall extensively use the following reduction
to indicate a simple way how to derive
the equations~(\ref{R1})-(\ref{R3}).
Let us do the shift $u_+\to u_++\lambda\ ,\ u_-\to u_-+\lambda\
,\ v_+\to v_++\lambda\ ,\ v_-\to v_-+\lambda$ in the
defining equation. The $\R$-operator is invariant under
this shift and the $\mathrm{L}$-operators transform in a simple way
$$
\mathrm{L}_1(u_++\lambda,u_-+\lambda) =
\left(\begin{array}{cc}
0 & \mathrm{S}^{-}_1 \\
z_1\cdot\left[z_1\dd_1+u_+-u_-\right]_q & 0
   \end{array}\right) + \frac{q^{\lambda}}{q-q^{-1}}\cdot
   \left(\begin{array}{cc}
q^{u_++z_1\dd_1} & 0 \\
0 & q^{u_--z_1\dd_1}
   \end{array}\right)+\{q\mapsto q^{-1}\}
$$
$$
\mathrm{L}_2(v_++\lambda,v_-+\lambda) =
\left(\begin{array}{cc}
0 & \mathrm{S}^{-}_2 \\
z_2\cdot\left[z_2\dd_2+v_+-v_-\right]_q & 0
   \end{array}\right) +
\frac{q^{\lambda}}{q-q^{-1}}\cdot
   \left(\begin{array}{cc}
q^{v_++z_2\dd_2} & 0 \\
0 & q^{v_--z_2\dd_2}
   \end{array}\right) +\{q\mapsto q^{-1}\}
$$
where the term $\{q\mapsto q^{-1}\}$ is obtained from the
second term by the substitution $q\mapsto q^{-1}$ so that
it is proportional to $q^{-\lambda}$. We obtain three sets
of equations as the consequence of the defining equation.
Matching coefficients in front of $q^{2\lambda}$ and
$q^{-2\lambda}$ we obtain the equation~(\ref{R1}). Matching
coefficients in front of $q^{\lambda}$ we derive~(\ref{R2})
and~(\ref{R3}) with sign plus and matching coefficients in
front of $q^{-\lambda}$ we obtain the same equations~($q
\mapsto q^{-1}$) with sign minus. It is possible to show
that the system of equations~(\ref{R1})-(\ref{R3}) is
equivalent to the initial defining
system~\cite{Jimbo,KKM,Bytsko}.

The general R-matrix can be represented as the product of
the simple building blocks denoted by $\RR^{\pm}$.
The main idea is very simple.
The operator $\check{\R}_{12}$ interchanges simultaneously
$u_{+}$ with $v_{+}$ and $u_{-}$ with $v_{-}$ in the product of two
Lax-operators
\be
\label{RLL}
\check{\R}_{12}(u_{+},u_-|v_{+},v_-)
\mathrm{L}_{1}(u_+,u_-)\mathrm{L}_{2}(v_+,v_-)=
\mathrm{L}_{1}(v_+,v_-)\mathrm{L}_{2}(u_+,u_-)
\check{\R}_{12}(u_{+},u_-|v_{+},v_-).
\ee
Let us perform this operation in two steps.
In the first step we interchange the parameters $u_{-}$ with
$v_{-}$ only. The parameters $u_{+}$ and $v_{+}$ remain the same.
In this way one obtains the natural defining equation for
the operator $\RR^{-}$.
In the case when $u_{-}=v_{-}=v$ there is no interchange of
parameters so that it is natural to expect that
the operator $\RR^{-}_{12}(u_{+},u_{-}|v_{-})$
is reduced to the unit operator $ \RR^{-}_{12}(u_{+},v|v) = \II $.

\begin{prop}
There exists the operator $\RR^{-}_{12}$ which is the solution of
the defining equations
\begin{equation}
\RR^{-}_{12} \mathrm{L}_{1}(u_+,u_-)\mathrm{L}_{2}(v_+,v_-)=
\mathrm{L}_{1}(u_+,v_-)\mathrm{L}_{2}(v_+,u_-) \RR^{-}_{12}
\label{R-}
\end{equation}
$$
\RR^-_{12} = \RR^-_{12}(u_+,u_-|v_-)\ ;\ \RR^-_{12}(u_+,u_-|v_-) =
\RR^-_{12}(u_++\lambda,u_-+\lambda|v_-+\lambda)
$$
The defining system~(\ref{R-}) is equivalent to the system
\begin{equation}
\RR^-_{12} \cdot q^{z_1\dd_1+z_2\dd_2} =
q^{z_1\dd_1+z_2\dd_2} \cdot \RR^{-}_{12}
\label{R-1}\ ;\ \RR^-_{12}\cdot z_2 = z_2\cdot\RR^-_{12}
\end{equation}
\begin{equation}
\RR^-_{12} \cdot \left(
\mathrm{S}^{-}_1\cdot
q^{\pm(v_--z_2\dd_2)} +
\mathrm{S}^{-}_2\cdot
q^{\pm(u_++z_1\dd_1)}\right) =
\left(
\mathrm{S}^{-}_1\cdot
q^{\pm(u_--z_2\dd_2)} +
\mathrm{S}^{-}_2\cdot
q^{\pm(u_++z_1\dd_1)}
\right)\cdot \RR^{-}_{12}
\label{R-2}
\end{equation}
$$
\RR^-_{12} \cdot \left(
z_1\cdot\left[z_1\dd_1+u_+-u_-\right]_q
\cdot q^{\pm(v_++z_2\dd_2)} +
z_2\cdot\left[z_2\dd_2+v_+-v_-\right]_q
\cdot q^{\pm(u_--z_1\dd_1)} \right) =
$$
\begin{equation}
= \left( z_1\cdot\left[z_1\dd_1+u_+-v_-\right]_q \cdot
q^{\pm(v_++z_2\dd_2)} +
z_2\cdot\left[z_2\dd_2+v_+-u_-\right]_q \cdot
q^{\pm(v_--z_1\dd_1)} \right)\cdot \RR^-_{12} \label{R-3}
\end{equation}
These conditions fix the operator $\RR^{-}_{12}$ up to
overall normalization constant.Fixing the normalization
in a such way that $\RR^{-}_{12}: 1\mapsto 1$ we obtain
\begin{equation}
\label{RR-}
\RR^{-}_{12}(u_+,u_-|v_-)=\frac{
\mathbf{e}_{q^2}\left(q^{2u_+-2u_-}\right)}{
\mathbf{e}_{q^2}\left(q^{2u_+-2v_-}\right)}\cdot
\mathbf{e}_{q^2}\left(q^{u_--u_++2}
\mathbf{u}_{12}\right) \cdot
\mathbf{v}_{1}^{\frac{v_--u_-}{2}}\cdot
\frac{
\mathbf{e}_{q^2}\left(q^{2u_+-2v_-}\mathbf{v}_{1}\right)}{
\mathbf{e}_{q^2}\left(q^{2u_+-2u_-}\mathbf{v}_{1}\right)} \cdot
\mathbf{e}^{-1}_{q^2}\left(q^{v_--u_++2}\mathbf{u}_{12}\right)
\end{equation}
where we use the operators forming a Weyl pair
\begin{equation}
\mathbf{u}_{12}\equiv \frac{z_2}{z_1}
\left[1-q^{-2z_1\dd_1}\right]\ ,\ \mathbf{v}_{1}\equiv q^{2z_1\dd_1}
\ ;\ \mathbf{u}_{12}\cdot\mathbf{v}_{1} =
q^2\cdot\mathbf{v}_{1}\cdot\mathbf{u}_{12}
\label{W1}
\end{equation}
\end{prop}

We use the standard q-exponential function~$\mathbf{e}_q
(x)$~\cite{Kir,FK,FV,Kashaev,V1}. The definition and useful
formulae with the function $\mathbf{e}_q (x)$ are collected
in Appendix.

In the second step we interchange $u_{+}$ with $v_{+}$ but
the parameters $u_{-}$ and $v_{-}$ remain the same.The
defining equation for the operator $\RR^{+}$ is
$$
\RR^{+}_{12}\cdot
\mathrm{L}_1(u_{+},u_{-})\mathrm{L}_2(v_{+},v_{-}) =
\mathrm{L}_1(v_{+},u_{-})\mathrm{L}_2(u_{+},v_{-})\cdot
\RR^{+}_{12} \ ;\ \RR^{+}_{12} =
\RR^{+}_{12}(u_{+}|v_{+},v_{-}).
$$
In the case when $u_{+}=v_{+}=u$ there should be the
similar degeneracy $\RR^{+}_{12}(u|u,v_{-}) = \II.$

\begin{prop}
There exists the operator $\RR^{+}_{12}$ which is the solution of
the defining equations
\begin{equation}
\RR^{+}_{12} \mathrm{L}_{1}(u_+,u_-)\mathrm{L}_{2}(v_+,v_-)=
\mathrm{L}_{1}(v_+,u_-)\mathrm{L}_{2}(u_+,v_-) \RR^{+}_{12}
\label{R+}
\end{equation}
$$
\RR^+_{12} = \RR^+_{12}(u_+|v_+,v_-)\ ;\
\RR^+_{12}(u_+|v_+,v_-) =
\RR^+_{12}(u_++\lambda|v_++\lambda,v_-+\lambda)
$$
The defining system~(\ref{R+}) is equivalent to the system
\begin{equation}
\RR^+_{12} \cdot q^{z_1\dd_1+z_2\dd_2} =
q^{z_1\dd_1+z_2\dd_2} \cdot \RR^{+}_{12}
\ ;\ \RR^+_{12}\cdot z_1 = z_1\cdot\RR^+_{12}
\label{R+1}
\end{equation}
\begin{equation}
\RR^+_{12} \cdot \left(
\mathrm{S}^{-}_1\cdot
q^{\pm(v_--z_2\dd_2)} +
\mathrm{S}^{-}_2\cdot
q^{\pm(u_++z_1\dd_1)}\right) =
\left(
\mathrm{S}^{-}_1\cdot
q^{\pm(v_--z_2\dd_2)} +
\mathrm{S}^{-}_2\cdot
q^{\pm(v_++z_1\dd_1)}
\right)\cdot \RR^{+}_{12}
\label{R+2}
\end{equation}
$$
\RR^+_{12} \cdot \left(
z_1\cdot\left[z_1\dd_1+u_+-u_-\right]_q
\cdot q^{\pm(v_++z_2\dd_2)} +
z_2\cdot\left[z_2\dd_2+v_+-v_-\right]_q
\cdot q^{\pm(u_--z_1\dd_1)} \right) =
$$
\begin{equation}
= \left(
z_1\cdot\left[z_1\dd_1+v_+-u_-\right]_q
\cdot q^{\pm(u_++z_2\dd_2)} +
z_2\cdot\left[z_2\dd_2+u_+-v_-\right]_q
\cdot q^{\pm(u_--z_1\dd_1)}
\right)\cdot \RR^+_{12}
\label{R+3}
\end{equation}
These conditions fix the operator $\RR^{+}_{12}$ up to
overall normalization constant.Fixing the normalization
in a such way that $\RR^{+}_{12}: 1\mapsto 1$ we obtain
\begin{equation}
\label{RR+} \RR^{+}_{12}(u_+|v_+,v_-) = \frac{
\mathbf{e}_{q^2}\left(q^{2v_+-2v_-}\right)}{
\mathbf{e}_{q^2}\left(q^{2u_+-2v_-}\right)} \cdot
\mathbf{e}_{q^2}\left(q^{v_--v_++2} \mathbf{u}_{21}\right)
\cdot \mathbf{v}_{2}^{\frac{v_+-u_+}{2}}\cdot \frac{
\mathbf{e}_{q^2}\left(q^{2u_+-2v_-}\mathbf{v}_{2}\right)}{
\mathbf{e}_{q^2}\left(q^{2v_+-2v_-}\mathbf{v}_{2}\right)}
\cdot
\mathbf{e}^{-1}_{q^2}\left(q^{2+v_--u_+}\mathbf{u}_{21}\right)
\end{equation}
\end{prop}

Note that the defining equations for the operator $\RR^{+}_{12}$
can be obtained from the defining equations for the
operator $\RR^{-}_{12}$ by simple change
$$
z_1 \leftrightarrow z_2\ ,\ u_- \leftrightarrow -v_+
\ ,\ u_+ \leftrightarrow -v_-
$$
so that the pair of plus-equations transforms to
the pair minus-equations and vice versa.
The second Weyl pair is
\begin{equation}
\mathbf{u}_{21}\equiv \frac{z_1}{z_2}
\left[1-q^{-2z_2\dd_2}\right]\ ,\ \mathbf{v}_{2}\equiv q^{2z_2\dd_2}
\ ;\ \mathbf{u}_{21}\cdot\mathbf{v}_{2} =
q^2\cdot\mathbf{v}_{2}\cdot\mathbf{u}_{21}
\label{W2}
\end{equation}

\begin{prop}
The operator $\check{\R}$ can be factorized in the following
way
\begin{equation}
\check{\R}_{12}(u_{+},u_-|v_{+},v_-)=
\RR^+_{12}(u_+|v_+,u_-)\RR^-_{12}(u_+,u_-|v_{-}) =
\RR^-_{12}(v_+,u_-|v_-)\RR^+_{12}(u_+|v_+,v_{-})
\label{Rfact}
\end{equation}
\end{prop}

The factorization of the $\check{\R}$-operator can be
proven using simple pictures. The operator $\check{\R}$
interchanges all parameters in the product of two
$\mathrm{L}$-operators. The operator $\RR_{-}$ interchanges
the parameters $u_-$ and $v_-$ only and the operator
$\RR_{+}$ interchanges the parameters $u_+$ and $v_+$.
Using the operator $\RR_{+}\RR_{-}$ it is possible to
interchange parameters $u_+,v_+$ and $u_-,v_-$ in two steps
so that we obtain the first equality in~(\ref{Rfact}) as
the condition of commutativity for the diagram

\vspace{5mm} \unitlength 0.8mm \linethickness{0.4pt}
\begin{picture}(130.00,35.00)
\put(50.00,10.00){\vector(1,1){15.00}}
\put(33.00,20.00){\makebox(0,0)[cc]{$\RR^{-}_{12}(u_{+},u_-|v_-)$}}
\put(85.00,25.00){\vector(1,-1){15.00}}
\put(117.00,20.00){\makebox(0,0)[cc]{$\RR^+_{12}(u_{+}|v_{+},u_-)$}}
\put(25.00,10.00){\makebox(0,0)[cc]{$\mathrm{L}_1(u_+,u_-)
\mathrm{L}_2(v_+,v_-)$}}
\put(75.00,30.00){\makebox(0,0)[cc]{$\mathrm{L}_1(u_+,v_-)
\mathrm{L}_2(v_+,u_-)$}}
\put(125.00,10.00){\makebox(0,0)[cc]{$\mathrm{L}_1(v_+,v_-)
\mathrm{L}_2(u_+,u_-)$}}
\put(75.00,5.00){\makebox(0,0)[cc]{$\check{\R}_{12}(u_{+},u_-|v_+,v_{-})$}}
\put(55.00,10.00){\vector(1,0){40.00}}
\end{picture}
\vspace{5mm}

It is possible to exchange the parameters in the opposite
order and derive in a such way the second
equality in~(\ref{Rfact}).

By definition  $\RR^{-}_{12}$ acts on the list of
parameters involved its defining RLL relation (\ref{R-}) as
\be \label{listpm1} (u_+,u_-,v_+,v_-) \rightarrow (u_+,v_-,
v_+, u_-) \ee which is equivalent to the following change
in terms of original spectral and representation
parameters, \be \label{listell1} [u, \ell_1, v, \ell_2]
\rightarrow [u -\xi, \ell_1+\xi, v +\xi, \ell_2-\xi], \ \ \
\xi = \frac{u_- - v_-}{2} . \ee On the other hand, as it
was discussed in the proof above, this defining relation
allows for an extra shift of $v_+$ by $\mu$: $v_+\to
v_++\mu$ beyond the global shift of $u_{\pm}, v_{\pm}$. The
$\mu$-translation symmetry is equivalent to the
commutativity   $\RR^{-}_{12} z_2 = z_2 \RR^{-}_{12}$.
Choosing now the particular shift $\mu = u_+ - v_+$, the
parameter changing action (\ref{listpm1}) becomes instead
\be \label{listpm2} (u_+,u_-,u_+,v_-) \rightarrow (u_+,v_-,
u_+, u_-) \ee
The RLL relation with these parameters in the
Lax matrices is just the one for the full $\check{\R}$
operator~(\ref{RLL}) and this confirms just that \be
 \RR^{-}_{12} (u_+,u_-|v_-) =
\check{\R}_{12} (u_+,u_-,u_+,v_-) \ee and shows also that
the known eigenvalue relations for $\check{\R}_{12}$ imply
the eigenvalue relations for $ \RR^{-}_{12} $. The former
are~\cite{KKM} \bea \label{Rphi} \check{\R}_{\ell_1,
\ell_2} (u-v) \ \ \varphi_n\left(z_1,z_2;\ell_1+
\frac{u-v}{2},\ell_2 + \frac{u-v}{2}\right)= \cr = \rho(n;
u-v,\ell_1,\ell_2) \ \ \varphi_n\left(z_1,z_2;\ell_2-
\frac{u-v}{2},\ell_1 - \frac{u-v}{2}\right) \eea with the
eigenfunctions \be \label{phin}
\varphi_n(z_1,z_2;\ell_1,\ell_2) = \prod_{k=1}^{n}
\left(q^{\ell_1 +k-1}  \ z_1 - q^{-\ell_2 -k +1} \ z_2
\right) = (-z_2)^n q^{-n\ell_2-\frac{n(n-1)}{2}}\cdot
\phi_n\left(\frac{z_1}{z_2}|\ell_1+\ell_2\right)\ee
$$
\phi_n\left(\frac{z_1}{z_2}|\ell_1+\ell_2\right)\equiv
\left(\frac{z_1}{z_2}q^{\ell_1+\ell_2};q^2\right)_{n}=
\frac{\mathbf{e}_{q^2}\left(\frac{z_1}{z_2}
q^{\ell_1+\ell_2+2n}\right)}
{\mathbf{e}_{q^2}\left(\frac{z_1}{z_2}q^{\ell_1+\ell_2}\right)}
$$
and the eigenvalues \be \label{eigenv} \rho (n; u-v,
\ell_1, \ell_2) = \prod_{k=1}^{n} {[u-v+ \ell_1 +\ell_2
+k-1]_q \over [v - u+ \ell_1 +\ell_2 +k-1]_q }. \ee Let us
rewrite these eigenvalue relations in terms of parameters
$u_{\pm}$ and $v_{\pm}$
$$
\check{\R}_{12} (u_+,u_-,v_+,v_-)(-z_2)^n
\phi_n\left(\frac{z_1}{z_2}|u_+-v_-\right) =
q^{n\left(u_--v_-\right)} \prod_{k=1}^{n} {[u_+-v_-+k-1]_q
\over [v_+-u_-+k-1]_q } \cdot
(-z_2)^n\phi_n\left(\frac{z_1}{z_2}|v_+-u_-\right)
$$
Next we put $v_+ = u_+$ and use the commutativity
$\RR^{-}_{12} z_2 = z_2 \RR^{-}_{12}$ to cancel the factor
$(-z_2)^n$.
\begin{prop}
The following generalized  eigenvalue relations hold with
the Yang-Baxter factor $\RR^{-}_{12}$ \be \RR^{-}_{12}
(u_+,u_-|v_-)\phi_n\left(\frac{z_1}{z_2}|u_+-v_-\right) =
q^{n\left(u_--v_-\right)} \prod_{k=1}^{n} {[u_+-v_-+k-1]_q
\over [u_+-u_-+k-1]_q } \cdot
\phi_n\left(\frac{z_1}{z_2}|u_+-u_-\right)
\label{R-phi}
\ee
\end{prop}

Now we are going to the proof of the Proposition 2. The
defining system of equations~(\ref{R-}) for the
$\RR^{-}$-operator can be reduced to the simpler system of
equations~(\ref{R-1})-(\ref{R-3}) similar the reduction for
the $\check{\R}_{12}$-operator. Let us make the shift
$u_+\to u_++\lambda\ ,\ u_-\to u_-+\lambda\ ,\ v_+\to
v_++\lambda+\mu\ ,\ v_-\to v_-+\lambda$ in the defining
equation~(\ref{R-}).

There is one difference to the
derivation in the case of operator $\check{\R}_{12}$:
the shift of parameter $v_+$ is independent and due to
the arbitrariness of $\mu$ we obtain the additional
condition of commutativity
$\RR^-_{12}\cdot z_2 = z_2\cdot\RR^-_{12}$.
It remains to solve the defining system of
plus-equations~(\ref{R-1})-(\ref{R-3}) and derive the explicit
expression for the operator $\RR^-_{12}$.
Using the commutation relations $\RR^{-}_{12} z_2 = z_2 \RR^{-}_{12}$
and $\RR^{-}_{12} q^{z_1\dd_1+z_2\dd_2}=
q^{z_1\dd_1+z_2\dd_2}\RR^{-}_{12}$ it
is possible to transform these two equations to the form
\begin{equation}
\RR^-_{12}\cdot \left(
1-q^{v_--u_+}\frac{z_2}{z_1}
\left[1-q^{-2z_1\dd_1}\right]\right)\cdot q^{2z_1\dd_1}
=
\left(1-q^{u_--u_+}\frac{z_2}{z_1}
\left[1-q^{-2z_1\dd_1}\right]\right)\cdot
q^{2z_1\dd_1}\cdot\RR^-_{12}
\label{1}
\end{equation}
$$
\RR^-_{12}\cdot\left( z_1\left[q^{u_+-u_-}-
q^{u_--u_+-2z_1\dd_1}\right]+ z_2
q^{u_--v_--2z_1\dd_1}\right)=
$$
\begin{equation}
= \left(
z_1\left[q^{u_+-v_-}- q^{v_--u_+-2z_1\dd_1}\right]+ z_2
q^{v_--u_--2z_1\dd_1} \right)\cdot\RR^-_{12} \label{2}
\end{equation}
Let us begin with the first equation and rewrite
it in the compact form
$$
\RR^-_{12}\cdot \left(1-q^{v_--u_+} \mathbf{u}\right)
\cdot  \mathbf{v} =
\left(1-q^{u_--u_+} \mathbf{u}\right)\cdot
 \mathbf{v}\cdot\RR^-_{12}
$$
We use notation~(\ref{W1}) suppressing the subscripts at
$\mathbf{u}_{12}$, $\mathbf{v}_1$ during this proof.
We substitute the operator $\RR^-_{12}$ as follows
$$
\RR^{-}_{12} = \mathbf{a}\left(\mathbf{u}\right)\cdot
\RR\cdot \mathbf{b}^{-1}\left(\mathbf{u}\right).
$$
Then the equation~(\ref{1}) has the form
$$
\RR \cdot \mathbf{b}^{-1}\left(\mathbf{u}\right)
\cdot\left(1-q^{v_--u_+} \mathbf{u}\right)
\cdot \mathbf{v}\cdot
\mathbf{b}\left(\mathbf{u}\right)
= \mathbf{a}^{-1}\left(\mathbf{u}\right)
\cdot\left(1-q^{u_--u_+} \mathbf{u}\right)
\cdot  \mathbf{v}\cdot
\mathbf{a}\left(\mathbf{u}\right)\cdot\RR
$$
Next we move the operator $\mathbf{v}$ to the right using
commutation relation $\mathbf{u}\mathbf{v}=q^2\mathbf{v}\mathbf{u}$
$$
\RR \cdot \underline{\mathbf{b}^{-1}\left(\mathbf{u}\right)
\cdot\left(1-q^{v_--u_+} \mathbf{u}\right)\cdot
\mathbf{b}\left(q^{-2}\mathbf{u}\right)}
\cdot\mathbf{v} =
\underline{\mathbf{a}^{-1}\left(\mathbf{u}\right)
\cdot\left(1-q^{u_--u_+} \mathbf{u}\right)
\cdot\mathbf{a}\left(q^{-2}\mathbf{u}\right)}
\cdot\mathbf{v}\cdot\RR
$$
If we choose the functions $\mathbf{a}(\mathbf{u})$ and
$\mathbf{b}(\mathbf{u})$ obeying the natural recurrence relations
\begin{equation}
\mathbf{a}^{-1}(\mathbf{u})\left(1-q^{u_--u_+}\mathbf{u}\right)
\mathbf{a}(q^{-2}\mathbf{u}) = 1 \Longleftrightarrow
\mathbf{a}(q^2\mathbf{u}) =
(1-q^{u_--u_++2}\mathbf{u})\cdot
\mathbf{a}(\mathbf{u})
\label{reca}
\end{equation}
\begin{equation}
\mathbf{b}^{-1}(\mathbf{u})\left(1-q^{v_--u_+}\mathbf{u}\right)
\mathbf{b}(q^{-2}\mathbf{u}) = 1 \Longleftrightarrow
\mathbf{b}(q^2\mathbf{u}) =
(1-q^{v_--u_++2}\mathbf{u})\cdot
\mathbf{b}(\mathbf{u})
\label{recb}
\end{equation}
then the equation for the operator $\RR$
reduces to the simple commutativity condition
$\RR \cdot \mathbf{v} = \mathbf{v}\cdot\RR$.
The solutions of the recurrence relations are
$$
\mathbf{a}(\mathbf{u}) = \mathbf{e}_{q^2}
\left(q^{u_--u_++2}\mathbf{u}\right)
\ ;\ \mathbf{b}(\mathbf{u}) =
\mathbf{e}_{q^2}\left(q^{v_--u_++2}\mathbf{u}\right)
$$
The second equation~(\ref{2}) in terms of
operator $\RR$ has the form
$$
\RR\cdot\mathbf{b}^{-1}(\mathbf{u})\cdot\left(
z_1\left[q^{u_+-u_-}- q^{u_--u_+-2z_1\dd_1}\right]+ z_2
q^{u_--v_--2z_1\dd_1}\right)\cdot\mathbf{b}(\mathbf{u})=
$$
\begin{equation}
=\mathbf{a}^{-1}(\mathbf{u})\cdot\left(
z_1\left[q^{u_+-v_-}- q^{v_--u_+-2z_1\dd_1}\right]+
z_2 q^{v_--u_--2z_1\dd_1} \right)\cdot
\mathbf{a}(\mathbf{u})\cdot\RR .
\label{eq2}
\end{equation}
Both sides contain expression of the type
\begin{equation}
\label{expr}
\mathbf{e}^{-1}_{q^2}
\left[q^{\alpha+2}\frac{z_2}{z_1}\left(1-q^{-2z_1\dd_1}\right)
\right]
\cdot\left(
z_1\left[q^{\beta}- q^{-\beta-2z_1\dd_1}\right]+ z_2
q^{-\alpha-\beta-2z_1\dd_1}\right)\cdot
\mathbf{e}_{q^2}
\left[q^{\alpha+2}\frac{z_2}{z_1}\left(1-q^{-2z_1\dd_1}\right)\right]
\end{equation}
which can be transformed to the much simpler form. We
obtain the factorization
$$
\mathbf{e}_{q^2}
\left[q^{\alpha+2}\frac{z_2}{z_1}\left(1-q^{-2z_1\dd_1}\right)\right]
=
\mathbf{e}_{q^2}\left[q^{\alpha+2}\frac{z_2}{z_1}\right]\cdot
\mathbf{e}_{q^2}
\left[-q^{\alpha+2}\frac{z_2}{z_1}q^{-2z_1\dd_1}\right]
$$
using the general formula~(see Appendix)
$$
\mathbf{e}_{q^2}\left(\mathbf{U}+\mathbf{V}\right)=
\mathbf{e}_{q^2}\left(\mathbf{V}\right)\cdot
\mathbf{e}_{q^2}\left(\mathbf{U}\right) \ ;\
\mathbf{U}\mathbf{V} = q^2\cdot\mathbf{V}\mathbf{U};\
\mathbf{U} = -q^{\alpha+2}\cdot\frac{z_2}{z_1}\cdot
q^{-2z_1\dd_1} \ ;\ \mathbf{V} =
q^{\alpha+2}\cdot\frac{z_2}{z_1}
$$
and then it is possible to transform~(\ref{expr}) in two
steps. First we have
$$
\mathbf{e}^{-1}_{q^2}
\left[q^{\alpha+2}\frac{z_2}{z_1}\right]
\cdot\left(
z_1\left[q^{\beta}- q^{-\beta-2z_1\dd_1}\right]+ z_2
q^{-\alpha-\beta-2z_1\dd_1}\right)\cdot
\mathbf{e}_{q^2}
\left[q^{\alpha+2}\frac{z_2}{z_1}\right]=
$$
$$
= z_1\cdot q^\beta -q^{-\beta}
\left(z_1-z_2\cdot q^{-\alpha}\right)\cdot
\mathbf{e}^{-1}_{q^2}
\left[q^{\alpha+2}\frac{z_2}{z_1}\right]
\cdot
\mathbf{e}_{q^2}
\left[q^{\alpha+2}\frac{z_2}{z_1}\cdot q^2\right]\cdot q^{-2z_1\dd_1}
=
$$
$$
= z_1\cdot q^\beta -
q^{-\beta}\cdot\left(z_1-z_2\cdot q^{-\alpha}\right)
\left(1-q^{\alpha+2}\frac{z_2}{z_1}\right)\cdot q^{-2z_1\dd_1}
$$
and in the next step we transform in a similar way the rest part
$$
\mathbf{e}_{q^2}^{-1}
\left[-q^{\alpha+2}\frac{z_2}{z_1}q^{-2z_1\dd_1}\right]
\left\{z_1\cdot q^\beta -
q^{-\beta}\cdot\left(z_1-z_2\cdot q^{-\alpha}\right)
\left(1-q^{\alpha+2}\frac{z_2}{z_1}\right)\cdot q^{-2z_1\dd_1}
\right\}
\mathbf{e}_{q^2}
\left[-q^{\alpha+2}\frac{z_2}{z_1}q^{-2z_1\dd_1}\right]=
$$
$$
= z_1\left(q^{\beta}-q^{-\beta-2z_1\dd_1}\right)-
q^{2-\beta}\frac{z^2_2}{z_1}
\left(1-q^{2\alpha+2-2z_1\dd_1}\right)
\left(1-q^{-2z_1\dd_1}\right)q^{-2z_1\dd_1}+
$$
$$
+z_2\left(q^{\alpha+\beta}-q^{-\alpha-\beta}\right)q^{-2z_1\dd_1}
+z_2q^{\alpha-\beta}
\left(q^{-2-2z_1\dd_1}-q^{2-2z_1\dd_1}-q^{-4z_1\dd_1}\right)
$$
After substitution of the particular expressions for the parameters
$\alpha$ and $\beta$ it appears that the expression
in the last line is the same for the left and right
hand sides of the equation~(\ref{eq2}). Due to the commutation relations
$\left[\RR ,q^{-2z_1\dd_1}\right] =
\left[\RR , z_2\right]= 0$ it is possible to cancel these terms
and we obtain
$$
\RR\cdot\left\{z_1q^{u_+-u_-}\left(1-q^{2u_--2u_+-2z_1\dd_1}\right)-
q^{2+v_--u_+}\frac{z^2_2}{z_1}
\left(1-q^{2+2u_--2u_+-2z_1\dd_1}\right)
\left(1-q^{-2z_1\dd_1}\right)q^{-2z_1\dd_1}\right\}=
$$
$$
= \left\{z_1q^{u_+-v_-}\left(1-q^{2v_--2u_+-2z_1\dd_1}\right)-
q^{2+u_--u_+}\frac{z^2_2}{z_1}
\left(1-q^{2+2v_--2u_+-2z_1\dd_1}\right)
\left(1-q^{-2z_1\dd_1}\right)q^{-2z_1\dd_1}\right\}\cdot \RR
$$
Since $z_2$ commutes with $\RR$ this condition implies two
relations , which however turn out to be equivalent to
$$
\RR\cdot z_1 q^{u_+-u_-}\left(1-q^{2u_--2u_+-2z_1\dd_1}\right)
= z_1 q^{u_+-v_-}\left(1-q^{2v_--2u_+-2z_1\dd_1}\right)\cdot\RR
$$
The solution can be represented in the form
$\RR = \RR(\mathbf{v})$ where $\mathbf{v} = q^{2z_1\dd_1}$ and
we obtain the recurrence relation
$$
\RR(q^2 \mathbf{v}) = q^{v_--u_-}
\frac{1-q^{2u_+-2v_-}\mathbf{v}}
{1-q^{2u_+-2u_-}\mathbf{v}}
\cdot\RR(\mathbf{v})
$$
The solution is
$$
\RR(\mathbf{v}) = \frac{\mathbf{e}_{q^2}\left(q^{2u_+-2u_-}\right)}
{\mathbf{e}_{q^2}\left(q^{2u_+-2v_-}\right)}
\cdot\mathbf{v}^{\frac{v_--u_-}{2}}\cdot
\frac{
\mathbf{e}_{q^2}\left(q^{2u_+-2v_-}\mathbf{v}\right)}{
\mathbf{e}_{q^2}\left(q^{2u_+-2u_-}\mathbf{v}\right)}
$$
where we fix the initial condition as follows $\RR(1) = 1$.

Collecting everything together one obtains the expression
for the operator $\RR^-_{12}$ in the form~(\ref{RR-}).

In the above proof we have used the conditions~(\ref{R-1})-(\ref{R-3})
with the upper sign plus only. The corresponding conditions
with the lower sign minus are obtained by the exchange $q\mapsto q^{-1}$.
In the Appendix we proof that
the result~(\ref{RR-}) can be written
more compactly.

\begin{prop}
The above operator result~(\ref{RR-}) can be written
 compactly as
\begin{equation}
\label{R-1c}
\RR^{-}_{12} =
\frac{\mathbf{e}_{q^2}\left(q^{2u_+-2u_-}\right)}
{\mathbf{e}_{q^2}\left(q^{2u_+-2v_-}\right)}\cdot
\mathbf{e}^{-1}_{q^2}\left(q^{2u_+-2u_-}\mathbf{v}_1
+q^{u_+-u_-}\bar{\mathbf{u}}_{12}\right) \cdot
\mathbf{v}_1^{\frac{v_--u_-}{2}}\cdot
\mathbf{e}_{q^2}\left(q^{2u_+-2v_-}\mathbf{v}_1
+q^{u_+-v_-}\bar{\mathbf{u}}_{12}\right)=
\end{equation}
\begin{equation}
\label{RR-2c}
=\frac{\mathbf{e}_{q^2}\left(q^{2v_--2u_++2}\right)}
{\mathbf{e}_{q^2}\left(q^{2u_--2u_++2}\right)}\cdot
\mathbf{e}_{q^2}\left(q^{2u_--2u_++2}\bar{\mathbf{v}}_1
+q^{u_--u_++2}\mathbf{u}_{12}\right) \cdot
\bar{\mathbf{v}}_1^{\frac{u_--v_-}{2}}\cdot
\mathbf{e}^{-1}_{q^2}\left(q^{2v_--2u_++2}\bar{\mathbf{v}}_1
+q^{v_--u_++2}\mathbf{u}_{12}\right)
\end{equation}
\end{prop}
Here $\bar{\mathbf{v}}_1$ and $\bar{\mathbf{u}}_{12}$ are obtained from
$\mathbf{v}_1$ , $\mathbf{u}_{12}$~(\ref{W1})
by replacing $q \mapsto q^{-1}$.
These expressions show the invariance of $\RR^{-}$
under $q \mapsto q^{-1}$ expected from the defining conditions


Using this compact operator representation it is
 instructive to rederive the result on eigenfunctions
 in Proposition 5
by calculating   directly the action of $ \RR^{-}_{12} $ on
functions of the above type (\ref{phin}).

The operator arguments of the functions $ \mathbf{e}_{q^2}
$ in (\ref{R-1c}) are \be \label{Abeta} A(\beta) =
q^{2\beta}\mathbf{v}_1 + q^{\beta}\bar{\mathbf{u}}_{12} =
q^{2\beta+2z_1\dd_1}+ q^{\beta}\frac{z_2}{z_1}
\left(1-q^{2z_1\dd_1}\right) \ee with $\beta$ substituted
by $u_+-v_-$ or $u_+ -u_-$. We have
$$
\phi_n\left(\frac{z_1}{z_2}|\beta\right)\equiv
\frac{\mathbf{e}_{q^2}
\left(\frac{z_1}{z_2}q^{\beta+2n}\right)}
{\mathbf{e}_{q^2}\left(\frac{z_1}{z_2}q^{\beta}\right)} \
;\ \phi_n\left(\frac{z_1}{z_2}\cdot q^2|\beta\right) =
\frac{1-\frac{z_1}{z_2}q^{\beta+2n}}
{1-\frac{z_1}{z_2}q^{\beta}}\cdot
\phi_n\left(\frac{z_1}{z_2}|\beta\right)
$$
$$
A(\beta)\cdot \phi_n\left(\frac{z_1}{z_2}|\beta\right)=
q^{2\beta} \phi_n\left(\frac{z_1}{z_2}\cdot
q^2|\beta\right) + q^{\beta}\frac{z_2}{z_1}\cdot \left[
\phi_n\left(\frac{z_1}{z_2}|\beta\right)
-\phi_n\left(\frac{z_1}{z_2}\cdot q^2|\beta\right)\right]=
$$
$$
=\phi_n\left(\frac{z_1}{z_2}|\beta\right)\cdot
\left\{q^{2\beta}\frac{1-\frac{z_1}{z_2}q^{\beta+2n}}
{1-\frac{z_1}{z_2}q^{\beta}} + q^{\beta}\frac{z_2}{z_1}
\left[1-\frac{1-\frac{z_1}{z_2}q^{\beta+2n}}
{1-\frac{z_1}{z_2}q^{\beta}}\right]\right\}=
q^{2\beta+2n}\cdot \phi_n\left(\frac{z_1}{z_2}\cdot
q^2|\beta\right)
$$
Therefore we have the eigenvalue relation for all $n$, \be
\label{Abetag} A(\beta)
\phi_n\left(\frac{z_1}{z_2}|\beta\right) = q^{2n+2\beta}
\phi_n\left(\frac{z_1}{z_2}|\beta\right).
\ee

This implies for (\ref{R-1c}) \be \label{R-z} \RR^{-} \ = {
\mathbf{e}_{q^2}\left(q^{2u_+-2u_-}\right) \over
\mathbf{e}_{q^2}\left(q^{2u_+ - 2v_-} \right) } \cdot
\mathbf{e}_{q^2}^{-1}\left(A(u_+-u_-) \right) \ \ q^{(u_- -
v_-) z_1 \dd_1} \ \ \mathbf{e}_{q^2}\left(A(u_+-v_-)
\right) \ee the eigenvalue relation \bea \label{R-g}
\RR^{-} \ \phi_n\left(\frac{z_1}{z_2}|u_+-v_-\right) =
{\mathbf{e}_{q^2}\left(q^{2u_+-2u_-}\right) \over
\mathbf{e}_{q^2}\left(q^{2u_+-2v_-}\right)} \cdot {
\mathbf{e}_{q^2}\left(q^{2n + 2(u_+-v_-)} \right) \over
\mathbf{e}_{q^2} \left(q^{2n + 2 (u_+-u_-)} \right) } \ \
\phi_n\left(\frac{z_1}{z_2}|u_+-u_-\right) \eea

Finally it is easy to check the coincidence of the
eigenvalues in (\ref{R-phi}) and (\ref{R-g}).

The compact form of $\RR^{-} $ given in Proposition 6 is
the direct q-deformed analogon of the corresponding result
obtained in \cite{SD2} for the undeformed case. The analogy
is also evident in the expressions for the eigenvalues and
the eigenfunctions \cite{KKM}.  The limit $q\rightarrow 1$,
discussed shortly in the Appendix, is \be \label{qto1}
\RR^{-} \to \RR^{-}_{XXX} =
\frac{\Gamma(u_+-u_-)}{\Gamma(u_+-v_-)}
\frac{\Gamma(z_{12}\dd_1+u_+-v_-)}{\Gamma(z_{12}\dd_1+u_+-u_-)}.
\ee


\section{ $\mathbf{Q}$-operators for
the generic  periodic XXZ spin chain.}

\setcounter{equation}{0}

The next natural step it to use the general operator
$\R_{12}(u_{+},u_{-}|v_{+},v_{-})$ as building block in
construction of Baxter Q-operators. In the case of the
generic inhomogeneous periodic XXZ spin chain the transfer
matrix $\mathbf{t}(u)$ is constructed as follows
$$
\mathbf{t}(u)=
\tr\mathrm{L}_1(u+\delta_1)\cdot\mathrm{L}_1(u+\delta_2)
\cdots\mathrm{L}_N(u+\delta_N)\ ;\ \mathrm{L}_k(u+\delta_k)
= \left(\begin{array}{cc}
\left[u+\delta_k+\mathrm{S}_k\right]_q &
\mathrm{S}^{-}_k \\
\mathrm{S}^{+}_k & \left[u+\delta_k-\mathrm{S}_k\right]_q
   \end{array}\right).
$$
The most general transfer matrix $\mathbf{Q}(u,\ell_{0})$
is constructed in a similar manner from the Yang Baxter operators
$\R_{k0}$
\be \label{Qtr}
\mathbf{Q}(u,\ell_{0}) =
\tr_{\mathrm{V}_{0}}\R_{10}(u+\delta_1)\R_{20}(u+\delta_2)
\cdots\R_{N0}(u+\delta_N)
\ee
The operator $\mathbf{Q}(u,\ell_{0})$ depends on two
parameters: the spectral parameter $u$ and the spin in the
auxiliary space $\mathrm{V}_{0}=\mathrm{V}_{\ell_0}$. It is
useful to change to other parameters $u_1=1+u-\ell_0$ and
$u_2 = u+\ell_{0}$ such that $\mathbf{Q}(u,\ell_{0})=
\mathbf{Q}(u_1|u_2)$.The explicit expression for the
operator $\R_{k0}$ is
$$
\R_{k0}(u+\delta_k) = \P_{k0}\cdot \frac{
\mathbf{e}_{q^2}\left(q^{-2u-2\delta_k+2\ell_k+2\ell_0}\right)}
{\mathbf{e}_{q^2}\left(q^{2u+2\delta_k
+2\ell_k+2\ell_0}\right)} \cdot
$$
$$
\cdot\mathbf{e}_{q^2}\left(q^{2+u+\delta_k-\ell_k-\ell_0}
\mathbf{u}_{0k}\right) \cdot
\mathbf{v}_{0}^{\frac{\ell_0-u-\delta_k-\ell_k}{2}}\cdot
\frac{
\mathbf{e}_{q^2}\left(q^{4\ell_k}\mathbf{v}_{0}\right)}{
\mathbf{e}_{q^2}\left(q^{2\ell_0+2\ell_k-2u-2\delta_k}
\mathbf{v}_{0}\right)} \cdot
\mathbf{e}^{-1}_{q^2}\left(q^{2-2\ell_k}\mathbf{u}_{0k}\right)
\cdot
$$
$$
\cdot \mathbf{e}_{q^2}\left(q^{2-2\ell_k}
\mathbf{u}_{k0}\right) \cdot
\mathbf{v}_{k}^{\frac{\ell_k-\ell_0-u-\delta_k}{2}}\cdot
\frac{ \mathbf{e}_{q^2}\left(q^{2u+2\delta_k
+2\ell_k+2\ell_0}\mathbf{v}_{k}\right)}{
\mathbf{e}_{q^2}\left(q^{4\ell_k}\mathbf{v}_{k}\right)}
\cdot
\mathbf{e}^{-1}_{q^2}\left(q^{2-u-\delta_k-\ell_k-\ell_0}
\mathbf{u}_{k0}\right)
$$
where
$$
\mathbf{u}_{0k}\equiv \frac{z_k}{z_0}
\left[1-q^{-2z_0\dd_0}\right]\ ,\ \mathbf{v}_{0}\equiv
q^{2z_0\dd_0}\ ;\ \mathbf{u}_{k0}\equiv \frac{z_0}{z_k}
\left[1-q^{-2z_k\dd_k}\right]\ ,\ \mathbf{v}_{k}\equiv
q^{2z_k\dd_k}.
$$
It is natural to simplify the notations: we omit the local
parameters $\ell_k,\delta_k$ in the chain and display the
global parameters $u_1=1+u-\ell_0$ and $u_2 = u+\ell_{0}$
only
$$
\R_{k0}(u_1|u_2) = \P_{k0}\cdot \frac{
\mathbf{e}_{q^2}\left(q^{2-2u_1-2\delta_k+2\ell_k}\right)}
{\mathbf{e}_{q^2}\left(q^{2u_2+2\delta_k +2\ell_k}\right)}
\cdot
$$
$$
\cdot\mathbf{e}_{q^2}\left(q^{1+u_1+\delta_k-\ell_k}
\mathbf{u}_{0k}\right) \cdot
\mathbf{v}_{0}^{\frac{1-u_1-\delta_k-\ell_k}{2}}\cdot
\frac{
\mathbf{e}_{q^2}\left(q^{4\ell_k}\mathbf{v}_{0}\right)}{
\mathbf{e}_{q^2}\left(q^{2+2\ell_k-2u_1-2\delta_k}
\mathbf{v}_{0}\right)} \cdot
\mathbf{e}^{-1}_{q^2}\left(q^{2-2\ell_k}\mathbf{u}_{0k}\right)
\cdot
$$
\be \label{Rglob}
\cdot \mathbf{e}_{q^2}\left(q^{2-2\ell_k}
\mathbf{u}_{k0}\right) \cdot
\mathbf{v}_{k}^{\frac{\ell_k-u_2-\delta_k}{2}}\cdot \frac{
\mathbf{e}_{q^2}\left(q^{2u_2+2\delta_k
+2\ell_k}\mathbf{v}_{k}\right)}{
\mathbf{e}_{q^2}\left(q^{4\ell_k}\mathbf{v}_{k}\right)}
\cdot \mathbf{e}^{-1}_{q^2}\left(q^{2-u_2-\delta_k-\ell_k}
\mathbf{u}_{k0}\right).
\ee

In our context we define the Baxter operators
 $\mathbf{Q}(u_1|u_2)$ as operators acting
 in $\otimes_{k=1}^{N} V_{\ell_k}$
with the basic properties
\begin{itemize}
\item commutativity
\be \label{Qdef1}
\mathbf{Q}(u_1|u_2)\cdot\mathbf{Q}(v_1|v_2)=
\mathbf{Q}(v_1|v_2)\cdot\mathbf{Q}(u_1|u_2) \  ; \
\mathbf{Q}(u_1|u_2)\cdot\mathbf{t}(v)=
\mathbf{t}(v)\cdot\mathbf{Q}(u_1|u_2)
\ee
\item  Baxter relation  with respect to $u_2$
\be \label{Qdef2}
\mathbf{Q}(u_1|u)\cdot\mathbf{t}(u) =
\Delta_{+}(u)\mathbf{Q}(u_1|u+1)+
\Delta_{-}(u)\mathbf{Q}(u_1|u-1) \ ;\
\Delta_{\pm}(u)=[u+\delta_1\pm\ell_1]_q\cdots
[u+\delta_N\pm\ell_N]_q
\ee
\item Baxter relation with respect to $u_1$
\be \label{Qdef3}
\mathbf{t}(u)\cdot\mathbf{Q}(u|u_2) =
\frac{\Delta_{+}(u-1)\Delta_{-}(u)}{\Delta_{-}(u-1)}
\mathbf{Q}(u-1|u_2) + \Delta_{-}(u)\mathbf{Q}(u+1|u_2)
\ee
\end{itemize}
These properties allow to consider the operators
$\mathbf{Q}(u_1|u_2)$ as two-parametric Baxter Q-operators
\cite{Baxter,Skl2}.\\
\\
$\mathbf{Theorem}$ {\it The generalized transfer matrices
$\mathbf{Q} (u; \ell_0)$ (\ref{Qtr}) of the generic
periodic XXZ chain are Baxter operators
$\mathbf{Q}(u_1|u_2) $ in the sense of the above definition
with $u_1=1+u-\ell_0$\\ and $u_2 = u+\ell_{0}$.}

The commutativity $ \left[\mathbf{Q}(u_1|u_2),
\mathbf{Q}(v_1|v_2)\right]= 0 $ follows from the
Yang-Baxter equation for the general R-matrix
$$
\R_{00^{\prime}}(u-v)\R_{k0}(u)\R_{k0^{\prime}}(v) =
\R_{k0^{\prime}}(u)\R_{k0}(u)\R_{00^{\prime}}(u-v)
$$
where $\mathrm{V}_0=\mathrm{V}_{\ell_0}$ and
$\mathrm{V}_{0^{\prime}}=\mathrm{V}_{\ell_{0^{\prime}}}$
are two auxiliary spaces and
$\mathrm{V}_k=\mathrm{V}_{\ell_k}$ is the k-th quantum
space.

The commutativity $ \left[\mathbf{Q}(u_1|u_2),
\mathbf{t}(v)\right] = 0 $ follows from the special case of
the general Yang-Baxter relation
$$
\R_{k0}(u-v)\mathrm{L}_k(u)\mathrm{L}_0(v) =
\mathrm{L}_0(v)\mathrm{L}_k(u)\R_{k0}(u-v).
$$
All these formulae are standard and well known. The really
nontrivial part of the proof is the derivation of the Baxter relations.
It is the consequence of the important properties of the
$\RR$-operators - the following triangularity relations.
\begin{prop}
The following triangularity relations hold for the
operators $\RR^{-}_{12}\left(u_{+},u_{-}|0\right)$ and
$\RR^{+}_{12}\left(u_{+}|1,u_{-}\right)$
\begin{equation}
\mathbf{M}^{-1}_1\cdot\RR^{-}_{12}\left(u_{+},u_{-}|0\right)\cdot
\mathrm{L}_1\left(u_{+},u_{-}\right) \cdot\mathbf{M}_2 =
\label{trianglR-}
\end{equation}
$$
=\left(\begin{array}{cc}
[u_{+}]_q\cdot\RR^{-}_{12}\left(u_{+}+1,u_{-}+1|0\right) &
-\RR^{-}_{12}\left(u_{+},u_{-}|0\right)\frac{1}{z_1}[z_1\dd_1]_q\\
0 &
[u_{-}]_q\cdot\RR^{-}_{12}\left(u_{+}-1,u_{-}-1|0\right)
\end{array}\right),\
$$
\begin{equation}
\mathbf{M}^{-1}_1\cdot
\mathrm{L}_2\left(u_{+},u_{-}\right)\cdot
\RR^{+}_{12}\left(u_{+}|1,u_{-}\right) \cdot\mathbf{M}_2 =
\label{trianglR+}
\end{equation}
$$
= \left(\begin{array}{cc} \frac{[u_-]_q[u_+ -1]_q}{[u_-
-1]_q}\cdot\RR^{+}_{12}\left(u_{+}-1|1,u_{-}-1\right) &
-\frac{1}{z_1}[z_1\dd_1]_q
\RR^{+}_{12}\left(u_{+}|1,u_{-}\right)\\
0 & [u_-]_q\cdot\RR^{+}_{12}\left(u_{+}+1|1,u_{-}+1\right)
\end{array}\right),\
$$
where
$$
\RR^{-}_{12}(u_+,u_-|0) = \frac{
\mathbf{e}_{q^2}\left(q^{2u_+-2u_-}\right)}{
\mathbf{e}_{q^2}\left(q^{2u_+}\right)}\cdot
\mathbf{e}_{q^2}\left(q^{u_--u_++2} \mathbf{u}_{12}\right)
\cdot \mathbf{v}_{1}^{-\frac{u_-}{2}}\cdot \frac{
\mathbf{e}_{q^2}\left(q^{2u_+}\mathbf{v}_{1}\right)}{
\mathbf{e}_{q^2}\left(q^{2u_+-2u_-}\mathbf{v}_{1}\right)}
\cdot
\mathbf{e}^{-1}_{q^2}\left(q^{2-u_+}\mathbf{u}_{12}\right)
$$
$$
\RR^{+}_{12}(u_{+}|1,u_-) = \frac{
\mathbf{e}_{q^2}\left(q^{2-2u_-}\right)}{
\mathbf{e}_{q^2}\left(q^{2u_+-2u_-}\right)} \cdot
\mathbf{e}_{q^2}\left(q^{u_-+1} \mathbf{u}_{21}\right)
\cdot \mathbf{v}_{2}^{\frac{1-u_+}{2}}\cdot \frac{
\mathbf{e}_{q^2}\left(q^{2u_+-2u_-}\mathbf{v}_{2}\right)}{
\mathbf{e}_{q^2}\left(q^{2-2u_-}\mathbf{v}_{2}\right)}
\cdot
\mathbf{e}^{-1}_{q^2}\left(q^{2+u_--u_+}\mathbf{u}_{21}\right)
$$
$$
\mathbf{M}_k \equiv \left(\begin{array}{cc}
1 & 0 \\
z_k& 1\end{array}\right).
$$
\end{prop}

We prove the triangularity relation for the operator
$\RR^{-}_{12}$ and the proof for the operator
$\RR^{+}_{12}$ is very similar. We start directly from the
defining equation~(\ref{R-}). Using
factorization~(\ref{factor}) of the Lax operator and
commutativity of $\RR^{-}_{12}$ and $z_2$ the defining
equation for the $\RR^{-}_{12}$-operator can be represented
in the form
$$
\RR^{-}_{12} \mathrm{L}_1(u_+,u_-)
\left(\begin{array}{cc}
1& 1 \\
z_{2}\cdot q^{-v_-}& z_{2}\cdot q^{v_-}
   \end{array}\right)
\left(\begin{array}{cc}
q^{z_2\dd_2}& 0 \\
0 & q^{-z_2\dd_2}
   \end{array}\right)
= \left(\begin{array}{cc}
1 & 1 \\
z_1\cdot q^{-v_-}& z_{1}\cdot q^{v_-}\end{array}\right)
\cdot
$$
$$
\cdot\left(\begin{array}{cc}
q^{z_1\dd_1}& 0 \\
0 & q^{-z_1\dd_1}
   \end{array}\right)
\left(\begin{array}{cc}
q^{u_+} & -\frac{q}{z_1} \\
-q^{-u_+} & \frac{1}{q z_1}\end{array}\right)\
\left(\begin{array}{cc}
1 & 1 \\
z_{2}\cdot q^{-u_-}& z_{2}\cdot q^{u_-}
   \end{array}\right)
\left(\begin{array}{cc}
q^{z_2\dd_2}& 0 \\
0 & q^{-z_2\dd_2}
   \end{array}\right)
\RR^{-}_{12}
$$
Next we transform all this in the following simple way
$$
\left(\begin{array}{cc}
1 & 0 \\
-z_1& 1\end{array}\right)\ \RR^{-}_{12}
\mathrm{L}_1(u_+,u_-)
\left(\begin{array}{cc}
1& 0 \\
z_{2}& 1
   \end{array}\right)=
$$
$$
=\left(q^{v_-}-q^{-v_-}\right)^{-1}\cdot
\left(\begin{array}{cc}
1 & 1 \\
z_1\cdot \left(q^{-v_-}-1\right)& z_{1}\cdot
\left(q^{v_-}-1\right)\end{array}\right)\
\left(\begin{array}{cc}
q^{z_1\dd_1}& 0 \\
0 & q^{-z_1\dd_1}
   \end{array}\right)
\left(\begin{array}{cc}
q^{u_+} & -\frac{q}{z_1} \\
-q^{-u_+} & \frac{1}{q z_1}\end{array}\right)\cdot
$$
$$
\cdot\left(\begin{array}{cc}
1 & 1 \\
z_{2}\cdot q^{-u_-}& z_{2}\cdot q^{u_-}
   \end{array}\right)
\left(\begin{array}{cc}
q^{z_2\dd_2}& 0 \\
0 & q^{-z_2\dd_2}
   \end{array}\right)\cdot
\RR^{-}_{12}\cdot \left(\begin{array}{cc}
q^{-z_2\dd_2}& 0 \\
0 & q^{z_2\dd_2}
   \end{array}\right)
\left(\begin{array}{cc} q^{-v_-}-1 & -\frac{1}{z_1}
\\ 1-q^{-v_-} & \frac{1}{z_2}
\end{array}\right)
$$
It is evident from this expression that matrix element in
the second row and first column is
$\sim\left(1-q^{v_-}\right)$ so that in the limit $v_- = 0$
we obtain the upper triangular matrix. To be sure, the
explicit calculations leads to the following expression for
this matrix element $\mathbf{A}_{21}$
$$
\mathbf{A}_{21} = \left(1-q^{v_-}\right)\cdot
\left[z_1\RR^{-}_{12}
\left(q^{u_+-v_-+z_1\dd_1}+q^{u_+-v_--z_1\dd_1}\right)
-z_2\RR^{-}_{12}
\left(q^{-u_-+z_1\dd_1}+q^{u_--v_--z_1\dd_1}\right)\right]
$$
and indeed we obtain zero at the point $v_-=0$. The
explicit calculation of the matrix elements
$\mathbf{A}_{11}, \mathbf{A}_{12}$ and $\mathbf{A}_{22}$ at
the point $v_- = 0$ gives the expression~(\ref{trianglR-}).
From the triangularity relations for the $\RR$-operators
immediately follow two triangularity relations for the
general operator $\R_{12}$.

\begin{prop}
The following triangularity relations for the operator
$\R_{12}\left(u_{+},u_{-}|v_+,v_-\right)$ hold
\begin{equation}
\mathbf{M}^{-1}_2\cdot \R_{12}\left(u_{+},u_{-}|v_+
,0\right)\cdot \mathrm{L}_1\left(u_{+},u_{-}\right)
\cdot\mathbf{M}_2 = \label{trianglR1}
\end{equation}
$$
=\left(\begin{array}{cc} [u_{+}]_q\cdot
\RR_{12}\left(u_{+}+1,u_{-}+1|v_+ +1, 0\right) &
*** \\
0 & [u_{-}]_q\cdot\R_{12}\left(u_{+}-1,u_{-}-1|v_+ -1,
0\right)
\end{array}\right)\
$$
\begin{equation}
\mathbf{M}^{-1}_2\cdot
\mathrm{L}_1\left(u_{+},u_{-}\right)\cdot
\R_{12}\left(u_{+},u_-|1,v_{-}\right) \cdot\mathbf{M}_2 =
\label{trianglR2}
\end{equation}
$$
= \left(\begin{array}{cc} \frac{[u_-]_q[u_+ -1]_q}{[u_-
-1]_q}\cdot\R_{12}\left(u_{+}-1,u_- -1|1,v_{-}-1\right) &
***\\
0 & u_-\cdot\R_{12}\left(u_{+}+1,u_- +1|1,v_{-}+1\right)
\end{array}\right)\
$$
\end{prop}

The relation~(\ref{trianglR1}) is obtained from the
relation~(\ref{trianglR-}) simply by multiplying with the
operator $\P_{12}\RR^{+}_{12}(u_+|v_+, u_-)$ from the left
and using the expression~(\ref{Rfact}) for the operator
$\R_{12}$. The operator $\RR^+$ plays a passive role in
this first relation. The relation~(\ref{trianglR2}) is
obtained from the relation~(\ref{trianglR+}) simply by
multiplying with the operator $\P_{12}$ from the left and
with the operator $\RR^{-}_{12}(u_+,u_-|v_-)$ from the
right and using the expression~(\ref{Rfact}) for the
operator $\R_{12}$. Now the operator $\RR^-$ plays a
passive role.

Let us go to the proof of the Baxter relation
$$
\mathbf{Q}(u_1|u)\cdot\mathbf{t}(u) =
\Delta_{+}(u)\mathbf{Q}(u_1|u+1)+
\Delta_{-}(u)\mathbf{Q}(u_1|u-1) \ ;\
\Delta_{\pm}(u)=[u+\delta_1\pm\ell_1]_q\cdots
[u+\delta_N\pm\ell_N]_q
$$
It is the direct consequence of the triangularity
relation~(\ref{trianglR1}) and cyclicity of the trace. Let
us choose the first space in~(\ref{trianglR1}) as k-th
quantum space and the second space as the auxiliary space.
We have in useful notations
$$
\mathbf{M}^{-1}_0\cdot\R_{k0}\left(u_1-v|u\right)\cdot
\mathrm{L}_k\left(u+\delta_k\right) \cdot\mathbf{M}_0 =
\left(\begin{array}{cc} [u^{+}_k]_q\cdot
\R_{k0}\left(u_1-v|u+1\right) &
*** \\
0 & [u^{-}_k]_q\cdot\R_{k0}\left(u_1-v|u-1\right)
\end{array}\right)\
$$
Multiplying these equalities for $k=1, 2, \cdots N $ ,
taking the traces in auxiliary spaces $\C^2$ and
$\mathrm{V}_0$ and using the cyclicity of the trace one
obtains the equation
$$
\mathbf{Q}(u_1-v|u)\cdot\mathbf{t}(u) =
\Delta_{+}(u)\mathbf{Q}(u_1-v|u+1)+
\Delta_{-}(u)\mathbf{Q}(u_1-v|u-1)
$$
The parameter $u_1$ is arbitrary so that we obtain the
needed relation. The Baxter relation with respect to
parameter $u_2$ follows from the triangularity
relation~(\ref{trianglR2}) and the derivation is very
similar.

\section{Q-operators for the homogeneous periodic XXZ spin chain}

\setcounter{equation}{0}

The operator $\R_{k0}(u_1|u_2)$~(\ref{Rglob}) has two
points of degeneracy: $u_1 = 1-\delta_k-\ell_k$ and $u_2 =
\ell_k-\delta_k$. In the case of homogeneous spin chain:
$\delta_k=0$ and $\ell_k=\ell$, the degeneration points for
all operators $\R_{k0}$ coincide so that it is possible to
remove half of the $\RR$-operators in the two-parametric
operator
$$
\mathbf{Q}(u_1|u_2) =
\tr_{\mathrm{V}_{0}}\R_{10}(u_1|u_2)\R_{20}(u_1|u_2)
\cdots\R_{N0}(u_1|u_2)
$$
We obtain the following reductions of the two-parametric
Q-operator: at the first point of degeneracy $u_1=1-\ell$
$$
\mathbf{Q}_{-}(u)=\mathbf{Q}(1-\ell|u)=
\tr_{\mathrm{V}_{0}}
\P_{10}\RR^{-}_{10}(u_{+},u_{-}|0)\cdot
\P_{20}\RR^{-}_{20}(u_{+},u_{-}|0)\cdots
\P_{N0}\RR^{-}_{N0}(u_{+},u_{-}|0)
$$
$$
\RR^{-}_{k0}(u_+,u_-|0) = \frac{
\mathbf{e}_{q^2}\left(q^{4\ell}\right)}{
\mathbf{e}_{q^2}\left(q^{2u+2\ell}\right)}\cdot
\mathbf{e}_{q^2}\left(q^{2-2\ell} \mathbf{u}_{k0}\right)
\cdot \mathbf{v}_{k}^{\frac{\ell-u}{2}}\cdot \frac{
\mathbf{e}_{q^2}\left(q^{2u+2\ell}\mathbf{v}_{k}\right)}{
\mathbf{e}_{q^2}\left(q^{4\ell}\mathbf{v}_{k}\right)} \cdot
\mathbf{e}^{-1}_{q^2}\left(q^{2-u-\ell}\mathbf{u}_{k0}\right)
$$
and at the second point of degeneracy $u_2=\ell$
$$
\mathbf{Q}_{+}(u)=\mathbf{Q}(u|\ell)= \tr_{\mathrm{V}_{0}}
\P_{10}\RR^{+}_{10}(u_{+}|1,u_{-})\cdot
\P_{20}\RR^{+}_{20}(u_{+}|1,u_{-})\cdots
\P_{N0}\RR^{+}_{N0}(u_{+}|1,u_{-})
$$
$$
\RR^{+}_{k0}(u_{+}|1,u_-) = \frac{
\mathbf{e}_{q^2}\left(q^{2+2\ell-2u}\right)}{
\mathbf{e}_{q^2}\left(q^{4\ell}\right)} \cdot
\mathbf{e}_{q^2}\left(q^{u+1-\ell} \mathbf{u}_{0k}\right)
\cdot \mathbf{v}_{2}^{\frac{1-\ell-u}{2}}\cdot \frac{
\mathbf{e}_{q^2}\left(q^{4\ell}\mathbf{v}_{0}\right)}{
\mathbf{e}_{q^2}\left(q^{2+2\ell-2u}\mathbf{v}_{0}\right)}
\cdot
\mathbf{e}^{-1}_{q^2}\left(q^{2-2\ell}\mathbf{u}_{0k}\right)
$$

As the direct consequence of the equations for the general
two-parametric operator $\mathbf{Q}(u_1|u_2)$ we
immediately derive the following properties of the
operators $\mathbf{Q}_{+}(u)$ and $\mathbf{Q}_{-}(u)$

\begin{prop}
For the homogeneous periodic XXZ chain the reductions
$\mathbf{Q}_{+}(u)$ and $\mathbf{Q}_{-}(u)$ arising from
the Baxter operators $\mathbf{Q}(u_1|u_2)$ at the points of
degeneracy $(u_1|u_2) = (1-\ell |u)$ and $(u_1|u_2) =
(u|\ell)$ obey the following relations:
\begin{itemize}
\item commutativity
\be \label{Qred1}
\mathbf{Q}_{\pm}(u)\cdot\mathbf{Q}_{\pm}(v)
=\mathbf{Q}_{\pm}(v)\cdot\mathbf{Q}_{\pm}(u) \ ;\
\mathbf{Q}_{+}(u)\cdot\mathbf{Q}_{-}(v)
=\mathbf{Q}_{-}(v)\cdot\mathbf{Q}_{+}(u) \  ; \
\mathbf{Q}_{\pm}(u)\cdot\mathbf{t}(v)=
\mathbf{t}(v)\cdot\mathbf{Q}_{\pm}(u)
\ee
\item Baxter relation for the $\mathbf{Q}_{-}(u)$
\be \label{Qred-}
\mathbf{Q}_{-}(u)\cdot\mathbf{t}(u) =
\Delta_{+}(u)\mathbf{Q}_{-}(u+1)+
\Delta_{-}(u)\mathbf{Q}_{-}(u-1) \ ;\
\Delta_{\pm}(u)=[u\pm\ell]_q^N
\ee
\item Baxter relation for the $\mathbf{Q}_{+}(u)$
\be \label{Qred+}
\mathbf{t}(u)\cdot\mathbf{Q_{+}}(u) =
\frac{\Delta_{+}(u-1)\Delta_{-}(u)}{\Delta_{-}(u-1)}
\mathbf{Q}_{+}(u-1) + \Delta_{-}(u)\mathbf{Q}_{+}(u+1).
\ee
\end{itemize}
\end{prop}
We construct the $\mathbf{Q}_{\pm}$-operator as the trace
of the products of $\RR^{\pm}$ operators in auxiliary space
$V_0$. The whole construction is pure algebraic. In this
Section we shall derive the explicit formulae for the
action of the operator $\mathbf{Q}_{-}$ in the space of
polynomials. The explicit expression for the second
operator $\mathbf{Q}_{+}$ is more complicated and we shall
not consider it here.

\begin{prop}
The action of the operator $\mathbf{Q}_{-}(u)$ on a
polynomial $\Psi(z_1\cdots z_N)$ can be represented in the
following form
\begin{equation}
\left[\mathbf{Q}_{-}(u)\Psi\right](z_1,\cdots z_N)= \left.
\RR_{01}\RR_{12}\RR_{23}\cdots
\RR_{N-1,N}\Psi(z_0,z_1,\cdots z_{N-1}) \right|_{z_0=z_N}
\label{QL1}
\end{equation}
where
$$
\RR_{k,k+1} = \frac{
\mathbf{e}_{q^2}\left(q^{4\ell}\right)}{
\mathbf{e}_{q^2}\left(q^{2u+2\ell}\right)}\cdot
\mathbf{e}_{q^2}\left(q^{2-2\ell} \mathbf{u}_{k,k+1}\right)
\cdot \mathbf{v}_{k}^{\frac{\ell-u}{2}}\cdot \frac{
\mathbf{e}_{q^2}\left(q^{2u+2\ell}\mathbf{v}_{k}\right)}{
\mathbf{e}_{q^2}\left(q^{4\ell}\mathbf{v}_{k}\right)} \cdot
\mathbf{e}^{-1}_{q^2}\left(q^{2-u-\ell}\mathbf{u}_{k,k+1}\right)
$$
$$
\mathbf{v}_{k} = q^{2z_k\dd_k} \ ,\ \mathbf{u}_{k,k+1} =
\frac{z_{k+1}}{z_k}\left[1-q^{-2z_k\dd_k}\right]
$$
\end{prop}
Let $z_0$ be the variable in the auxiliary space $V_0$ and
let the operator $\mathbb{A}$ act in the tensor product
$\mathrm{V}_{0}\otimes\mathrm{V}_{1}\cdots\otimes
\mathrm{V}_{N}$ and $\Psi(z_1\cdots z_N) \in
\mathrm{V}_{1}\cdots\otimes \mathrm{V}_{N}$. The trace of
the operator $\mathbb{A}$ in auxiliary space
$\mathrm{V}_{0}=\C[z_0]$ can be calculated as follows
$$
\left[\left(\tr_{\mathrm{V}_{0}} \mathbb{A}\right)
\Psi\right](z_1\cdots z_N) = \left.\sum_{m=0}^{+\infty}
\frac{1}{m!} \dd_0^m \mathbb{A} \cdot z_0^m
\cdot\Psi(z_1\cdots z_N)\right|_{z_0=0}.
$$
In order to prove~(\ref{QL1}) it is useful to move all
permutations to the right
$$ \P_{10}\RR_{10}\P_{20}\RR_{20}\cdots
\P_{N0}\RR_{N0} = \RR_{01}\RR_{12}\RR_{23}\cdots
\RR_{N-1,N} \cdot \P_{10}\P_{20}\cdots\P_{N0}.
$$
Then we have
$$
\RR_{01}\RR_{12}\RR_{23}\cdots \RR_{N-1,N}\cdot
\P_{10}\P_{20}\cdots\P_{N0}\cdot z_0^m \cdot\Psi(z_1\cdots
z_N) =
$$
$$
=\RR_{01}\RR_{12}\RR_{23}\cdots \RR_{N-1,N}\cdot z_N^m
\cdot\Psi(z_0,z_1\cdots z_{N-1})= z_N^m
\RR_{01}\RR_{12}\RR_{23}\cdots
\RR_{N-1,N}\Psi(z_0,z_1\cdots z_{N-1}).
$$
The result of the operation $\sum_{m=0}^{+\infty}
\frac{1}{m!} \dd_0^m$ can be calculated in closed form
$$
\left[\mathbf{Q}_{-}(u)\Psi\right](z_1\cdots z_N) =
\sum_{m=0}^{+\infty} \frac{1}{m!} \dd_0^m z_N^m
\left.\RR_{01}\RR_{12}\cdots \RR_{N-1,N}\Psi(z_0,z_1\cdots
z_{N-1})\right|_{z_0=0}=
$$
$$
= \mathrm{e}^{z_N\dd_0}
\left.\RR_{01}\RR_{12}\RR_{23}\cdots
\RR_{N-1,N}\Psi(z_0,z_1\cdots z_{N-1})\right|_{z_0=0}=
\left.\RR_{01}\RR_{12}\RR_{23}\cdots
\RR_{N-1,N}\Psi(z_0,z_1\cdots z_{N-1})\right|_{z_0=z_N}
$$
and we obtain the representation~(\ref{QL1}).

\section{Discussion}

\setcounter{equation}{0}

We have presented a factorization of the $R$ operator
acting on generic representations of the q-deformed $sl(2)$
algebra in close analogy to the undeformed case studied
earlier.

Starting from the Lax matrix with the q-deformed generators
represented as  functions of the Heisenberg pair of
operators $z, \dd $, the factors $\RR^{\pm}$ have been
derived in terms of functions of these basic operators. The
defining $\mathrm{R}\mathrm{L}\mathrm{L}$ relation for the
factors can be treated easier compared to the ones for the
full operators $\R$.

The basis of eigenfunctions of the $R$ matrix factors
$\RR^{\pm}$ has been derived both by relating the
eigenvalue problems to the one of the full operator $\R$
and by direct calculations.

The central results concern the construction of Baxter
operators $Q$ as generalized transfer matrices. The latter are
defined by the generic $\R$ operators replacing the Lax
matrices. The explicit form of the operators  $\R$ in terms
of their factors derived here allows for explicit
representations of the Baxter operators and the proof of
the Baxter relations. Triangularity properties based on the
factorization of the Lax matrix separating contributions
involving the operators $z$ and $\dd $ from each other turn
out to be the origin of the Baxter relations.

The results concern the case of periodic chains of XXZ
type. We do not expect difficulties in extensions to open
chains.

The special case of homogeneous periodic chains is of
particular interest. Here among the Baxter operators the
particular one where the representation parameter of the
auxiliary and the quantum spaces coincide ($\ell_0 = \ell$)
is distinguished; the nearest-neighbour interaction
hamiltonian can be constructed using this Q-operator.

Here other distinguished parameter values for Baxter
operators have been discussed, where the expressions as
generalized transfer matrices reduce to traces of the
factor operators $\RR^{\pm}$. The commutativity properties of the
resulting Baxter operators $Q_{\pm}$ have been given. We
have presented the explicit operator expression for $Q_-$
and its action on polynomial functions representing the
quantum states of the chain.

\section*{Acknowledgments}

S.D. would like to thank A. Bytsko for the stimulating
discussions and critical remarks. The work of S.D.
has been supported by the grant 03-01-00837 of the
Russian Foundation for Fundamental Research.
D.K. has been supported in part
by Volkswagen Stiftung  and by INTAS grant 03-51-5460.
We are grateful to Deutsche Forschungsgemeinschaft
for supporting our collaboration by a travel grant.
Two of us (S.D. and D.K.) are grateful to Leipzig University
and its Center of Theoretical Science for kind hospitality.

\section{Appendix}

\setcounter{equation}{0}

\subsection{The q-exponential function $\mathbf{e}_q (x)$}

In this Appendix we collect some useful
formulae~\cite{Kir,FK,FV,Kashaev,V1}. The standard
q-products are
$$
(x;q)_{\infty} = \prod_{k=0}^{+\infty}(1-q^k\cdot x)
\ ;\ (x;q)_{n} =  \prod_{k=0}^{n-1}(1-q^k\cdot x) =
\frac{(x;q)_{\infty}}{(xq^n;q)_{\infty}}
\ ;\ q\in \C \ ,\ |q| < 1
$$
The q-exponential function $\mathbf{e}_q (x)$ is
defined as follows
$$
\mathbf{e}_q (x) \equiv \frac{1}{(x;q)_{\infty}}
\ ;\ \mathbf{e}_q (q x) = (1-x)\cdot\mathbf{e}_q (x).
$$
The q-binomial formula
$$
\sum_{n=0}^{\infty}
\frac{(a;q)_{n}}{(q;q)_{n}}\cdot z^n =
\frac{(a z;q)_{\infty}}{(z;q)_{\infty}}\ ;\ |z| < 1
$$
allows to derive two expansion formulas
$$
\mathbf{e}_q (x) = \sum_{k=0}^{+\infty}\frac{x^n}{(q;q)_{n}}
\ ;\ \mathbf{e}^{-1}_q (x) =
\sum_{k=0}^{+\infty}\frac{(-1)^nq^{\frac{n(n-1)}{2}}x^n}{(q;q)_{n}}
$$
It is clear from these formulae that we can define a unique
extension of the function $\mathbf{e}_q (x)$
$$
\mathbf{e}_{q^{-1}}(x) =\mathbf{e}^{-1}_{q}(q x)
$$
so that the recurrence relation
$\mathbf{e}_q (q x) = (1-x)\cdot\mathbf{e}_q (x)$
still holds in large domain $|q|\neq 1$ .

There are important formulae
involving the Weyl pair $\mathbf{u}\mathbf{v} =
q\cdot\mathbf{v}\mathbf{u}$~\cite{FK}
$$
\mathbf{e}_q (\mathbf{v})\cdot\mathbf{e}_q (\mathbf{u}) =
\mathbf{e}_q (\mathbf{u}+\mathbf{v})
$$
$$
\mathbf{e}_q (\mathbf{u})\cdot\mathbf{e}_q (\mathbf{v}) =
\mathbf{e}_q (\mathbf{u}+\mathbf{v}-\mathbf{v}\mathbf{u})=
\mathbf{e}_q (\mathbf{v})\cdot
\mathbf{e}_q (-\mathbf{v}\mathbf{u})
\cdot\mathbf{e}_q (\mathbf{u})=
$$
$$
=
\mathbf{e}_q (\mathbf{v}-\mathbf{v}\mathbf{u})\cdot
\mathbf{e}_q (\mathbf{u})=
\mathbf{e}_q (\mathbf{v})\cdot
\mathbf{e}_q (\mathbf{u}-\mathbf{v}\mathbf{u})
$$

\subsection{Different representations for the operator $\RR^{-}_{12}$}

Let us introduce two Weyl pairs of operators which are connected
by the change $q \mapsto q^{-1}$
$$
\mathbf{u}\equiv \frac{z_2}{z_1}
\left[1-q^{-2z_1\dd_1}\right]\ ,\ \mathbf{v}\equiv q^{2z_1\dd_1}
\ ;\ \mathbf{u}\cdot\mathbf{v} =
q^2\cdot\mathbf{v}\cdot\mathbf{u}
$$
$$
\bar{\mathbf{u}}\equiv \frac{z_2}{z_1}
\left[1-q^{2z_1\dd_1}\right]\ ,\ \bar{\mathbf{v}}\equiv q^{-2z_1\dd_1}
\ ;\ \bar{\mathbf{u}}\cdot\bar{\mathbf{v}} =
q^{-2}\cdot\bar{\mathbf{v}}\cdot\bar{\mathbf{u}}
$$
There is the first pair of equivalent expressions for
the operator $\RR^{-}_{12}$
$$
\RR^{-}_{12} =
\frac{\mathbf{e}_{q^2}\left(q^{2u_+-2u_-}\right)}
{\mathbf{e}_{q^2}\left(q^{2u_+-2v_-}\right)}\cdot
\mathbf{e}_{q^2}\left(q^{u_--u_++2}
\mathbf{u}\right) \cdot
\mathbf{v}^{\frac{v_--u_-}{2}}\cdot
\frac{
\mathbf{e}_{q^2}\left(q^{2u_+-2v_-}\mathbf{v}\right)}{
\mathbf{e}_{q^2}\left(q^{2u_+-2u_-}\mathbf{v}\right)} \cdot
\mathbf{e}^{-1}_{q^2}\left(q^{v_--u_++2}\mathbf{u}\right)=
$$
\begin{equation}
\label{R-1lim}
= \frac{\mathbf{e}_{q^2}\left(q^{2u_+-2u_-}\right)}
{\mathbf{e}_{q^2}\left(q^{2u_+-2v_-}\right)}\cdot
\mathbf{e}^{-1}_{q^2}\left(q^{2u_+-2u_-}\mathbf{v}
+q^{u_+-u_-}\bar{\mathbf{u}}\right) \cdot
\mathbf{v}^{\frac{v_--u_-}{2}}\cdot
\mathbf{e}_{q^2}\left(q^{2u_+-2v_-}\mathbf{v}
+q^{u_+-v_-}\bar{\mathbf{u}}\right)
\end{equation}
The first expression is the one obtained in
the above proof of~(\ref{RR-}).
To derive the second expression from the first
one we transform the product
$\mathbf{e}_{q^2}\left(q^{u_--u_++2}
\mathbf{u}\right) \cdot
\mathbf{e}^{-1}_{q^2}\left(q^{2u_+-2u_-}\mathbf{v}\right)$
using the general formula
$$
\mathbf{e}_{q^2}\left(\mathbf{u}\right)
\mathbf{e}_{q^2}\left(\mathbf{v}\right)=
\mathbf{e}_{q^2}\left(\mathbf{v}-\mathbf{v}\mathbf{u}\right)
\mathbf{e}_{q^2}\left(\mathbf{u}\right)
\ ;\ \mathbf{u}\mathbf{u} = q^2 \mathbf{v}\mathbf{u}
$$
in the equivalent form
$$
\mathbf{e}_{q^2}\left(\mathbf{u}\right)
\mathbf{e}^{-1}_{q^2}\left(\mathbf{v}\right)
=\mathbf{e}^{-1}_{q^2}\left(\mathbf{v}-\mathbf{v}\mathbf{u}\right)
\mathbf{e}_{q^2}\left(\mathbf{u}\right)
$$
It is clear that operators $\lambda_1\cdot\mathbf{u}$ and
$\lambda_2\cdot\mathbf{v}$ also form Weyl pair so that
these general formulae are applicable.
Using this formula and equality
$-q^{2u_+-2u_-}\mathbf{v}\cdot q^{u_--u_++2}
\mathbf{u} = q^{u_+-u_-}\bar{\mathbf{u}}$ we derive
$$
\mathbf{e}_{q^2}\left(q^{u_--u_++2}
\mathbf{u}\right) \cdot
\mathbf{e}^{-1}_{q^2}\left(q^{2u_+-2u_-}\mathbf{v}\right)
=\mathbf{e}^{-1}_{q^2}\left(q^{2u_+-2u_-}\mathbf{v}
+q^{u_+-u_-}\bar{\mathbf{u}}\right) \cdot
\mathbf{e}_{q^2}\left(q^{u_--u_++2}
\mathbf{u}\right)
$$
The second product
$\mathbf{e}_{q^2}\left(q^{2u_+-2v_-}\mathbf{v}\right)
\cdot\mathbf{e}^{-1}_{q^2}\left(q^{v_--u_++2}\mathbf{u}\right)$
in the initial expression for $\RR^{-}_{12}$ is obtained
from the first one by inverse and change $u_-\mapsto v_-$
so that we have
$$
\mathbf{e}_{q^2}\left(q^{2u_+-2v_-}\mathbf{v}\right)
\cdot \mathbf{e}^{-1}_{q^2}\left(q^{v_--u_++2}\mathbf{u}\right)
= \mathbf{e}^{-1}_{q^2}\left(q^{v_--u_++2}\mathbf{u}\right)
\mathbf{e}_{q^2}\left(q^{2u_+-2v_-}\mathbf{v}
+q^{u_+-v_-}\bar{\mathbf{u}}\right)
$$
Finally two terms cancel each other
$$
\mathbf{e}_{q^2}\left(q^{u_--u_++2}
\mathbf{u}\right)\cdot
\mathbf{v}^{\frac{v_--u_-}{2}}
\cdot\mathbf{e}^{-1}_{q^2}\left(q^{v_--u_++2}\mathbf{u}\right)=
\mathbf{v}^{\frac{v_--u_-}{2}}
$$
and we derive the second relation.
There is the second pair of equivalent expressions for
the operator $\RR^{-}_{12}$
\begin{equation}
\label{RR-2}
\RR^{-}_{12}(u_+,u_-|v_-) =
\frac{
\mathbf{e}_{q^2}\left(q^{2v_--2u_++2}\right)}{
\mathbf{e}_{q^2}\left(q^{2u_--2u_++2}\right)}\cdot
\mathbf{e}^{-1}_{q^2}\left(q^{u_+-u_-}
\bar{\mathbf{u}}\right) \cdot
\bar{\mathbf{v}}^{\frac{u_--v_-}{2}}\cdot
\frac{
\mathbf{e}_{q^2}\left(q^{2u_--2u_++2}\bar{\mathbf{v}}\right)}{
\mathbf{e}_{q^2}\left(q^{2v_--2u_++2}\bar{\mathbf{v}}\right)} \cdot
\mathbf{e}_{q^2}\left(q^{u_+-v_-}\bar{\mathbf{u}}\right) =
\end{equation}
\begin{equation}
\label{RR-2lim}
=\frac{
\mathbf{e}_{q^2}\left(q^{2v_--2u_++2}\right)}{
\mathbf{e}_{q^2}\left(q^{2u_--2u_++2}\right)}\cdot
\mathbf{e}_{q^2}\left(q^{2u_--2u_++2}\bar{\mathbf{v}}
+q^{u_--u_++2}\mathbf{u}\right) \cdot
\bar{\mathbf{v}}^{\frac{u_--v_-}{2}}\cdot
\mathbf{e}^{-1}_{q^2}\left(q^{2v_--2u_++2}\bar{\mathbf{v}}
+q^{v_--u_++2}\mathbf{u}\right)
\end{equation}
The equation~(\ref{RR-2lim}) is derived from the
equation~(\ref{RR-2}) in the similar way as it
was done for the first pair.
To derive the equation~(\ref{RR-2}) we start from expression
$$
\RR^{-}_{12}(u_+,u_-|v_-)\sim
\mathbf{e}_{q^2}\left(q^{u_--u_++2}
\mathbf{u}\right) \cdot
\mathbf{v}^{\frac{v_--u_-}{2}}\cdot
\frac{
\mathbf{e}_{q^2}\left(q^{2u_+-2v_-}\mathbf{v}\right)}{
\mathbf{e}_{q^2}\left(q^{2u_+-2u_-}\mathbf{v}\right)} \cdot
\mathbf{e}^{-1}_{q^2}\left(q^{v_--u_++2}\mathbf{u}\right)
$$
and transform it using the relation
$$
\mathbf{e}_{q^2}\left(\mathbf{u}\right)
\mathbf{e}_{q^2}\left(\mathbf{v}\right)=
\mathbf{e}_{q^2}\left(\mathbf{v}\right)
\mathbf{e}_{q^2}\left(-\mathbf{v}\mathbf{u}\right)
\mathbf{e}_{q^2}\left(\mathbf{u}\right)
\ ;\ \mathbf{u}\mathbf{v} = q^2 \mathbf{v}\mathbf{u}
$$
in equivalent form
$$
\mathbf{e}_{q^2}\left(\mathbf{u}\right)
\mathbf{e}^{-1}_{q^2}\left(\mathbf{v}\right)=
\mathbf{e}^{-1}_{q^2}\left(-\mathbf{v}\mathbf{u}\right)
\mathbf{e}^{-1}_{q^2}\left(\mathbf{v}\right)
\mathbf{e}_{q^2}\left(\mathbf{u}\right)
$$
Using $-q^{2u_+-2u_-}\mathbf{v}\cdot q^{u_--u_++2}
\mathbf{u}=q^{u_+-u_-}\bar{\mathbf{u}}$ we have
$$
\mathbf{e}_{q^2}\left(q^{u_--u_++2}
\mathbf{u}\right)
\mathbf{e}^{-1}_{q^2}\left(q^{2u_+-2u_-}\mathbf{v}\right)
= \mathbf{e}^{-1}_{q^2}\left(q^{u_+-u_-}\bar{\mathbf{u}}\right)
\mathbf{e}^{-1}_{q^2}\left(q^{2u_+-2u_-}\mathbf{v}\right)
\mathbf{e}_{q^2}\left(q^{u_--u_++2}\mathbf{u}\right)
$$
and by inverse and change $u_- \mapsto v_-$
we derive
$$
\mathbf{e}_{q^2}\left(q^{2u_+-2v_-}
\mathbf{v}\right)
\mathbf{e}^{-1}_{q^2}\left(q^{v_--u_++2}\mathbf{u}\right)
= \mathbf{e}^{-1}_{q^2}\left(q^{v_--u_++2}\mathbf{u}\right)
\mathbf{e}_{q^2}\left(q^{2u_+-2v_-}
\mathbf{v}\right)
\mathbf{e}_{q^2}\left(q^{u_+-v_-+2}\bar{\mathbf{u}}\right)
$$
Again two term cancel each other
$$
\mathbf{e}_{q^2}\left(q^{u_--u_++2}\mathbf{u}\right)
\cdot
\mathbf{v}^{\frac{v_--u_-}{2}}
\cdot
\mathbf{e}^{-1}_{q^2}\left(q^{v_--u_++2}\mathbf{u}\right)=
\mathbf{v}^{\frac{v_--u_-}{2}}
$$
and one obtains the expression
$$
\mathbf{e}^{-1}_{q^2}\left(q^{u_+-u_-}\bar{\mathbf{u}}\right)
\underline{\mathbf{e}^{-1}_{q^2}\left(q^{2u_+-2u_-}\mathbf{v}\right)
\mathbf{v}^{\frac{v_--u_-}{2}}
\mathbf{e}_{q^2}\left(q^{2u_+-2v_-} \mathbf{v}\right)}
\mathbf{e}_{q^2}\left(q^{u_+-v_-+2}\bar{\mathbf{u}}\right).
$$
On the last step we transform the marked expression to the form
\begin{equation}
\mathbf{v}^{\frac{v_--u_-}{2}}\cdot
\frac{
\mathbf{e}_{q^2}\left(q^{2u_+-2v_-}\mathbf{v}\right)}{
\mathbf{e}_{q^2}\left(q^{2u_+-2u_-}\mathbf{v}\right)}
\cdot\frac{
\mathbf{e}_{q^2}\left(q^{2u_+-2u_-}\right)}{
\mathbf{e}_{q^2}\left(q^{2u_+-2v_-}\right)} =
\bar{\mathbf{v}}^{\frac{v_--u_-}{2}}\cdot
\frac{
\mathbf{e}_{q^2}\left(q^{2u_--2u_++2}\bar{\mathbf{v}}\right)}{
\mathbf{e}_{q^2}\left(q^{2v_--2u_++2}\bar{\mathbf{v}}\right)}
\cdot
\frac{
\mathbf{e}_{q^2}\left(q^{2v_--2u_++2}\right)}{
\mathbf{e}_{q^2}\left(q^{2u_--2u_++2}\right)}
\label{eq}
\end{equation}
and derive the expression~(\ref{RR-2}). The
equation~(\ref{eq}) is obtained as follows. Both sides of
this equation are solutions of the recurrence relation from
the above proof of~($\ref{RR-}$)
$$
\RR\cdot z_1 q^{u_+-u_-}\left(1-q^{2u_--2u_+-2z_1\dd_1}\right)
= z_1 q^{u_+-v_-}\left(1-q^{2v_--2u_+-2z_1\dd_1}\right)\cdot\RR
$$
$$
\RR(q^2 \mathbf{v}) = q^{v_--u_-}
\frac{1-q^{2u_+-2v_-}\mathbf{v}}
{1-q^{2u_+-2u_-}\mathbf{v}}
\cdot\RR(\mathbf{v})
\ ;\ \RR(q^2 \bar{\mathbf{v}}) = q^{v_--u_-}
\frac{1-q^{2u_--2u_++2}\bar{\mathbf{v}}}
{1-q^{2v_--2u_++2}\bar{\mathbf{v}}}
\cdot\RR(\bar{\mathbf{v}})
$$
The solutions have the form
$$
\RR(\mathbf{v}) = \mathbf{v}^{\frac{v_--u_-}{2}}\cdot
\frac{
\mathbf{e}_{q^2}\left(q^{2u_+-2v_-}\mathbf{v}\right)}{
\mathbf{e}_{q^2}\left(q^{2u_+-2u_-}\mathbf{v}\right)}
\cdot\frac{
\mathbf{e}_{q^2}\left(q^{2u_+-2u_-}\right)}{
\mathbf{e}_{q^2}\left(q^{2u_+-2v_-}\right)}
\ ;\ \RR(\bar{\mathbf{v}}) = \bar{\mathbf{v}}^{\frac{v_--u_-}{2}}\cdot
\frac{
\mathbf{e}_{q^2}\left(q^{2u_--2u_++2}\bar{\mathbf{v}}\right)}{
\mathbf{e}_{q^2}\left(q^{2v_--2u_++2}\bar{\mathbf{v}}\right)}
\cdot
\frac{
\mathbf{e}_{q^2}\left(q^{2v_--2u_++2}\right)}{
\mathbf{e}_{q^2}\left(q^{2u_--2u_++2}\right)}
$$
and equal initial conditions $\RR(1) = 1$ so that they coincide.

\subsection{The limit $q \to 1$ for the operators $\RR^{\pm}_{12}$}

Now we are going to show that in the case $q \to 1$ the
expression~(\ref{R-1lim})
$$
\RR^{-}_{12} =
\frac{\mathbf{e}_{q^2}\left(q^{2u_+-2u_-}\right)}
{\mathbf{e}_{q^2}\left(q^{2u_+-2v_-}\right)}\cdot
\mathbf{e}^{-1}_{q^2}\left(q^{2u_+-2u_-+2z_1\dd_{1}}
+q^{u_+-u_-}\frac{z_2}{z_1}\left(1-q^{2z_1\dd_1}\right)
\right) \cdot
$$
$$
\cdot q^{(v_--u_-)z_1\dd_1}\cdot
\mathbf{e}_{q^2}\left(q^{2u_+-2v_-+2z_1\dd_{1}}
+q^{u_+-v_-}\frac{z_2}{z_1}\left(1-q^{2z_1\dd_1}\right)
\right)
$$
reproduces the expression~\cite{SD2}
$$
\RR^{-}_{12} = \frac{\Gamma(u_+-u_-)}{\Gamma(u_+-v_-)}
\frac{\Gamma(z_{12}\dd_1+u_+-v_-)}{\Gamma(z_{12}\dd_1+u_+-u_-)}.
$$
Using the definition of the function
$$
\Gamma_q(z)=
\frac{(q;q)_{\infty}\cdot(1-q)^{1-z}}{(q^z;q)_{\infty}}=
\frac{\mathbf{e}_{q}(q^z)\cdot(1-q)^{1-z}}
{\mathbf{e}_{q}(q)}
$$
and the well known fact that $\Gamma_q(z) \to \Gamma(z)$
for $q \to 1$ we obtain the leading contribution for
$\epsilon \to 0$
$$
\frac{\mathbf{e}_{1-\epsilon}(1-\epsilon\cdot x )}
{\mathbf{e}_{1-\epsilon}(1-\epsilon\cdot y)} \to
\epsilon^{x-y}\cdot\frac{\Gamma(x)}{\Gamma(y)}.
$$
In the first order in $\epsilon$ we have~($q^2 \to
1-\epsilon$)
$$
q^{2 a} \to 1-\epsilon a\ ,\ q^{2b+2z_1\dd_{1}}+q^{b}
\frac{z_2}{z_1}\left(1-q^{2z_1\dd_1}\right) \to 1
-\epsilon\left(b+z_{12}\dd_1\right)
$$
so that
$$
\RR^{-}_{12} \to
\frac{\mathbf{e}_{1-\epsilon}\left(1-\epsilon(u_+-u_-)\right)}
{\mathbf{e}_{1-\epsilon}\left(1-\epsilon(u_+-v_-)\right)}\cdot
\frac{\mathbf{e}_{1-\epsilon} \left(1
-\epsilon\left(u_+-u_-+z_{12}\dd_1\right)\right)}
{\mathbf{e}_{1-\epsilon} \left(1
-\epsilon\left(u_+-v_-+z_{12}\dd_1\right)\right)} \to
\frac{\Gamma(u_+-u_-)}{\Gamma(u_+-v_-)}
\frac{\Gamma(z_{12}\dd_1+u_+-v_-)}{\Gamma(z_{12}\dd_1+u_+-u_-)}
$$
For the operator $\RR^{+}_{12}$ all calculations are
similar and in the limit $q \to 1$ one reproduces the
expression~\cite{SD2}
$$
\RR^{+}_{12} = \frac{\Gamma(v_+-v_-)}{\Gamma(u_+-v_-)}
\frac{\Gamma(z_{21}\dd_2+u_+-v_-)}{\Gamma(z_{21}\dd_2+v_+-v_-)}.
$$

\end{document}